\documentclass[12pt]{article}
\pdfoutput=1
\usepackage[]{hyperref}
\usepackage{xcolor}
\usepackage[]{putex}
\usepackage[utf8]{inputenc}

\usepackage{amsmath,amsthm,amssymb}
\usepackage{tikz}
\usepackage{tikz-cd}
\usetikzlibrary{decorations.markings}
\usetikzlibrary{calc}
\usepackage{subcaption}
\usepackage{enumerate}
\usepackage{enumitem}
\usepackage{changepage}

\definecolor{colour1}{rgb}{0, 0, 8/10}
\definecolor{colour2}{rgb}{0.4, 0.5, 1}
\definecolor{colour3}{rgb}{0.7, 0.85, 1}
\definecolor{colour4}{rgb}{0.922526, 0.385626, 0.209179}

\definecolor{colour5}{rgb}{0.560181, 0.9, 0.194885}
\definecolor{colour6}{rgb}{0.260181, 0.691569, 0.194885}
\definecolor{colour7}{rgb}{0.847624, 0.37816, 0.614037}
\definecolor{colour8}{rgb}{0.647624, 0.37816, 0.614037}

\title{Adelic Amplitudes
\\
and Intricacies of Infinite Products
}

\authors{Christian Baadsgaard Jepsen}
\institution{SCGP}{Simons Center for Geometry and Physics, Stony Brook University, Stony Brook, NY, 11794}

\abstract{
For every prime number $p$ it is possible to define a $p$-adic version of the Veneziano amplitude and its higher-point generalizations. Multiplying together the real amplitude with all its $p$-adic counterparts yields the adelic amplitude. At four points it has been argued that the adelic amplitude, after regulating the product that defines it, equals one. For the adelic 5-point amplitude, there exist kinematic regimes where no regularization is needed. This paper demonstrates that in special cases within this regime, the adelic product can be explicitly evaluated in terms of ratios of the Riemann zeta function, and observes that the 5-point adelic amplitude is not given by a single analytic function. Motivated by this fact to study new regularization procedures for the 4-point amplitude, an alternative formalism is presented, resulting in non-constant amplitudes that are piecewise analytic in the three scattering channels, including one non-constant adelic amplitude previously suggested in the literature. Decomposing the residues of these amplitudes into weighted sums of Gegenbauer polynomials, numerical evidence indicates that in special ranges of spacetime dimensions all the coefficients are positive, as required by unitarity.
}

\begin{document}

\maketitle
\tableofcontents

\section{Introduction}
String theory is a powerful and multifaceted formalism for studying topics in physics and pure math. Included among the subjects encompassed by the formalism are formulations of string theory based on the field $\mathbb{Q}_p$ of $p$-adic numbers. 

The notion of $p$-adic string theory was introduced independently by Volovich \cite{volovich1987p} and Grossman \cite{grossman1987p} in 1987, who each proposed a theory where spacetime and momenta are valued over the $p$-adic numbers $\mathbb{Q}_p$, and amplitudes are given in terms of Morita gamma functions. The conception of $p$-adic string theory which has since become standard, and which was put forward by Freund and Olson \cite{freund1987non}, operates with real-valued spacetime coordinates and momenta, while scattering amplitudes, now given by Gelfand-Graev gamma functions, are valued over the complex numbers; the role of the $p$-adic numbers is to serve as integration variables in the integral representation of $N$-point tachyon amplitudes $A_N^{(p)}$. As clarified by Zabrodin \cite{zabrodin1989non}, the worldsheet of $p$-adic string theory consists of a regular $(p+1)$-valent tree known as the Bruhat-Tits tree, while the boundary of the world sheet is given by $\mathbb{Q}_p$. The tachyon amplitudes obtained in this setup are significantly simpler than those of real bosonic string theory and contain only a finite number of poles, all due to tachyons, rather than semi-infinite sequences of poles arising from an infinite tower of higher spin particles. This simplicity has enabled $p$-adic string theory to serve as a helpful toy model, particularly in the context of string field theory. In 1988, long before the corresponding calculation was carried out in real string theory,  Brekke, Freund, Olson, and Witten \cite{brekke1988non} managed to compute the tree-level effective spacetime action for the tachyonic field of $p$-adic string theory. The stationary configurations of this action were subsequently interpreted by Ghoshal and Sen as exotic $D$-branes\footnote{ We refer the reader to Huang, Stoica, and Zhong's paper \cite{huang2021massless} for a recent proposal for a theory of $p$-adic 2-branes.} and understood to provide an explicit manifestation of tachyon condensation \cite{ghoshal2000tachyon}, an interpretation corroborated by Minahan's work \cite{minahan2001mode} demonstrating the consistency of tree-level $p$-adic string amplitudes and tachyon fluctuations about lumps in the effective action; but also an interpretation requiring modifications at loop-level \cite{minahan2001quantum}. Some 12 years after the computation of the spacetime effective action of $p$-adic string theory, Gerasimov and Shatashvili \cite{gerasimov2000exact} derived a tree-level Lagrangian for the tachyon of real open string theory. Their result allowed them to make the curious observation that their Lagrangian could have also been obtained by taking the $p\rightarrow 1$ limit of the tachyon Lagrangian in the $p$-adic formalism.

The idea that $p$-adic string theories, beyond serving as mere toy models, might enjoy a precise relation to real string theory was advanced already in 1987, when Freund and Witten \cite{freund1987adelic} raised the suggestion that all the $p$-adic string theories and real bosonic string theory might be commingled in a theory associated to the adelic numbers $\mathbb{A}$ that conjoin $\mathbb{R}$ with all the fields $\mathbb{Q}_p$ for every prime $p$. The first step towards explicating this suggestion is to present a unified description of 4-point amplitudes in real and $p$-adic string theory. To this end, it is customary to introduce a label $v$ such that $v=\infty$ or $v=p$ for some prime $p$, where the value $v=\infty$ refers to the real case, meaning that $\mathbb{Q}_\infty=\mathbb{R}$ and that $|\cdot|_\infty$ is the familiar absolute value norm, while $|\cdot|_p$ is the $p$-adic norm. In this notation, working with spacetime signature $(-,+,...,+)$, introducing dimensionless Mandelstam invariants given by
\begin{align}
s=\,-\alpha'(k_1+k_2)^2&\,, \nonumber
\\
t=\,-\alpha'(k_1+k_3)^2&\,, \label{Mandelstam}
\\
u=\,-\alpha'(k_1+k_4)^2&\,, \nonumber
\end{align}
and setting aside the question of overall normalization, the 4-point tachyon amplitudes in the various theories are given by
\begin{align}
A_4^{(v)}=\int_{\mathbb{Q}_v} dx\,|x|_v^{2\alpha' k_1\cdot k_2}|1-x|_v^{2\alpha' k_1\cdot k_3}
=\int_{\mathbb{Q}_v} dx\,|x|_v^{-2-s}|1-x|_v^{-2-t}\,.
\label{A4}
\end{align}
Throughout this paper, we will only be considering tree-amplitudes and only full amplitudes, not partial amplitudes. When $v=\infty$, we recover the standard Veneziano amplitude:
\begin{align}
A_4^{(\infty)}=
\frac{\Gamma(-1-s)\Gamma(-1-t)}{\Gamma(-2-s-t)}+
\frac{\Gamma(-1-s)\Gamma(-1-u)}{\Gamma(-2-s-u)}+
\frac{\Gamma(-1-t)\Gamma(-1-u)}{\Gamma(-2-t-u)}\,.
\label{A4real}
\end{align}
In the $p$-adic case, direct integration over $\mathbb{Q}_p$ gives
\begin{align}
A_4^{(p)}=&\,
\frac{\zeta_p(-1-s)\zeta_p(-1-t)\zeta_p(-1-u)}{\zeta_p(2+s)\zeta_p(2+t)\zeta_p(2+u)}
=\,\Gamma_p(-1-s)\Gamma_p(-1-t)\Gamma_p(-1-u)\,,
\label{A4padic}
\end{align}
where the local zeta and gamma functions are given by
\begin{align}
\zeta_p(x)=\frac{1}{1-p^{-x}}\,,\hspace{20mm}\Gamma_p(x)=\frac{\zeta_p(x)}{\zeta_p(1-x)}\,. \label{zetap}
\end{align}
There also exist related zeta and gamma functions associated to the real case $v=\infty$,
\begin{align}
\zeta_\infty(x)=\pi^{-\frac{x}{2}}\,\Gamma\big(\frac{x}{2}\big)\,,\hspace{20mm}\Gamma_\infty(x)=\frac{\zeta_\infty(x)}{\zeta_\infty(1-x)}\,,
\end{align}
and a fascinating observation of Freund and Witten's is that the three terms of the full Veneziano amplitude \eqref{A4real} combine into a single expression of the same form as the $p$-adic answer:
\begin{align}
A_4^{(\infty)}=&\,
\frac{\zeta_\infty(-1-s)\zeta_\infty(-1-t)\zeta_\infty(-1-u)}{\zeta_\infty(2+s)\zeta_\infty(2+t)\zeta_\infty(2+u)}
=\,\Gamma_\infty(-1-s)\Gamma_\infty(-1-t)\Gamma_\infty(-1-u)\,.
\end{align}
By invoking the functional equation for the Riemann zeta function
\begin{align}
\zeta_\infty(x)\,\zeta(x)=\zeta_\infty(1-x)\,\zeta(1-x)\,,
\end{align}
we can also express the Veneziano amplitude solely in terms of the Riemann zeta function:
\begin{align}
A_4^{(\infty)}=&\,
\frac{\zeta(2+s)\zeta(2+t)\zeta(2+u)}{\zeta(-1-s)\zeta(-1-t)\zeta(-1-u)}\,.
\label{VenezianoRiemann}
\end{align}
We now recall the formula due to Euler that expresses the Riemann zeta function as a product running over the set $\mathbb{P}$ of all primes:
\begin{align}
\prod_{p\in \mathbb{P}}\zeta_p(z) = \zeta(z)\,.
\label{Euler}
\end{align}
The product on the left-hand side is convergent when Re$[z]>1$, whereas the Riemann zeta function on the right-hand side admits an analytic continuation to the entire complex plane except for the simple pole at $z=0$. Following \cite{freund1987adelic}, we therefore observe that if we are allowed to distribute an infinite product of ratios of local zeta functions into each zeta function and to perform separate analytic continuations of these sub-products,
\begin{align}
\label{distribute}
\prod_p \frac{\zeta_p(...)\zeta_p(...)...}{\zeta_p(...)\zeta_p(...)...}
\,\leftrightarrow\,
 \frac{\Big(\prod_p\zeta_p(...)\Big)\Big(\prod_p\zeta_p(...)\Big)...}{\Big(\prod_p\zeta_p(...)\Big)\Big(\prod_p\zeta_p(...)\Big)...}
 \,\leftrightarrow\,
 \frac{\zeta(...)\zeta(...)...}{\zeta(...)\zeta(...)...}\,,
\end{align}
then we arrive at what is known as the adelic product formula for the Veneziano amplitude:
\begin{align}
A_4^{(\infty)}\prod_{p\in \mathbb{P}}A_4^{(p)}=1\,.
\label{A4adelic}
\end{align}
The assumption of distributivity \eqref{distribute}, or some other regularization scheme that results in the same effect, is necessary because there is no kinematic regime for which the product in equation \eqref{A4adelic} converges; the criterion Re$[z]>1$ required for the convergence of \eqref{Euler} cannot be simultaneously satisfied for all the local zeta functions in \eqref{A4padic}.\footnote{There are arguments in the literature \cite{vladimirov1993freund,huang2020green,huang2022quadratic} attesting to the validity of the equation
\begin{align}
\Gamma_\infty(z)\prod_{p\in\mathbb{P}}\Gamma_p(z)=1
\label{gammaProd}
\end{align}
when understood in a suitably regulated sense, but even with equation \eqref{gammaProd} given, an assumption of distributivity is needed in order to split the product over $A_4^{(p)}$ into three products over local gamma functions.}

These issues of non-convergence can be circumvented by turning to the adelic 5-point amplitude, for which there exist kinematic regimes where the prime product does converge. Introducing the shorthands
\begin{align}
s_{ij} \equiv 2\alpha'k_i \cdot k_j\,,
\hspace{20mm}
s_{ijk} \equiv 2\alpha'k_i\cdot k_j+2\alpha'k_i\cdot k_k+2\alpha'k_j\cdot k_k\,,
\end{align}
the real $(v=\infty)$ and $p$-adic $(v=p)$ tachyonic 5-point amplitudes are given by 
\cite{virasoro1969generalization,bardakci1969meson,koba1969reaction,fairlie1970integral}
\begin{align}
A_5^{(v)}=&
\int_{\mathbb{Q}_v} dx\,dy\,|x|_v^{s_{12}}\,|y|_v^{s_{13}}\,|y-x|_v^{s_{23}}\,
|1-y|_v^{s_{34}}\,|1-x|_v^{s_{24}}\,.
\label{A5}
\end{align}
In the real case, the double-integral can be carried out analytically as in \cite{kitazawa1987effective}, giving 
\begin{align}
& \hspace{40mm} 
A_5^{(\infty)} = \frac{1}{10} \sum_{\sigma \in S_5} F\big[\sigma(\vec{k}_1,...,\vec{k}_5)\big]\,,
\nonumber\\[-16pt]
\hspace{1mm}
\\ \nonumber
&\text{where} 
\hspace{5mm}
F\big[\vec{k}_1,...,\vec{k}_5\big]
=
B\big(s_{123}+2,s_{34}+1\big)B\big(s_{12}+1,s_{23}+1\big)
\\
&\hspace{43mm}
{}_3F_2
\Big[
\big\{
-s_{24},\,s_{12}+1,\,s_{123}+2
\big\},
\big\{
s_{123}+s_{34}+3,\,
s_{12}+s_{23}+2
\big\};
1
\Big]\,,
\nonumber
\end{align}
and the factor of $\frac{1}{10}$ cancels over-counting due to cyclic permutations and their inversions. For the $p$-adics, the five-point amplitude, as computed in \cite{marinari1988p,brekke1988non}, evaluates to\footnote{ We refer the reader to \cite{stoica2021closed} for a recent analysis of the parallelism between the exact evaluations of the real and $p$-adic 5-point amplitudes.}
\begin{align}
A_5^{(p)}=
\frac{(p-2)(p-3)}{p^2}
+
\sum_{i\neq j}\frac{(p-1)(p-2)}{2p^2(p^{1+s_{ij}}-1)}
+
\sum_{i,j,k,l\text{ distinct}}
\frac{(p-1)^2}{8p^2(p^{1+s_{ij}}-1)(p^{1+s_{kl}}-1)}\,.
\end{align}
While evaluating the adelic product $A_5^{(\infty)}\prod_{p\in\mathbb{P}}A_5^{(p)}$ presents a somewhat formidable task, Refs. \cite{marinari1988p,brekke1988non} were each able to argue, by demonstrating that the zeros and poles of $A_5^{(\infty)}$ and $\prod_{p\in\mathbb{P}}A_5^{(p)}$ do not align, that this product is not a momentum-independent constant and so in particular does not equal one.  As with the derivation of the adelic 4-point formula \eqref{A4adelic}, the arguments of both references require splitting an infinite product over $\mathbb{P}$ into several such products, which do not all converge. 

With the benefit modern computing power, it is not hard to check numerically that $A_5^{(\infty)}\prod_{p\in\mathbb{P}}A_5^{(p)}$ is indeed not a constant in the kinematic regions where the product converges. Furthermore, as will be explicitly demonstrated in the next section of this paper, there exists a codimension-one subset of kinematic space for which the adelic product $A_5^{(\infty)}\prod_{p\in\mathbb{P}}A_5^{(p)}$ can be evaluated exactly. The exact result reveals two important facts: 
the 5-point tree-amplitude is not an analytic function of the kinematic invariants, and it is not generally valid to apply the identifications \eqref{distribute} to the adelic 5-point product. Since analyticity and the distribution of an infinite product into subfactors played a key role in the regularization of the adelic four-point amplitude sketched above, we are motivated to investigate alternative regularization methods. Such an alternative was considered previously by Aref'eva, Dragovi\'c and Volovich \cite{aref1988adelic}, who observed that while the product $\prod_{p\in\mathbb{P}}A_4^{(p)}(s,t)$ never converges, there are $s$- and $t$-values for which the product $\prod_{p\in\mathbb{P}}\big(-A_4^{(p)}(s,t)\big)$ does converge and can be interpreted as the $p$-adic component of a non-constant adelic 4-point amplitude. This amplitude is given by a separate function for each scattering channel and therefore does not obey crossing symmetry, for which reason it has generated sparse interest. However, since we will discover in the next section that the 5-point adelic amplitude possesses precisely such a piecewise analytic structure, the proposal of Ref. \cite{aref1988adelic} merits closer inspection. Taking seriously the notion of a non-constant adelic 4-point amplitude, it behoves us to inquire whether such an amplitude complies with the demands of causality, locality, and unitarity, and also whether there are other candidates for non-constant adelic amplitudes besides that of \cite{aref1988adelic}. Finally, we may pose the question, what is the critical dimension of the adelic string?

To address these questions, the present paper adopts a prescription for regulating divergent products that does not rely on splitting an infinite product via $\prod_n\big(a_n \,b_n\big) \leftrightarrow \big(\prod_n a_n\big)\big(\prod_n b_n\big)$ and then performing a separate analytic continuation for each sub-product. In applying this prescription to adelic products, we arrive at a number of non-constant candidate amplitudes, which are given by ratios of Riemann zeta functions or other Dirichlet $L$-functions. To evaluate the suitability of these expressions as actual scattering amplitudes, we will analyze their partial wave expansions and study their high energy asymptotics. In so doing, we discover that while the candidate amplitudes exhibit unorthodox pole structures and only piecewise analyticity, they display the well-behaved high energy behaviour of conventional string amplitudes as well as, within particular ranges of target space dimensionality, the positivity properties required by unitarity.

\section{The Adelic 5-point Amplitude at Special Kinematics}
\label{2}

The real and $p$-adic 5-point amplitudes \eqref{A5} simplify greatly if we consider the special case when any two momenta are orthogonal. We will consider the case when $k_2\cdot k_3=0$, noting that for tachyonic scattering such a kinematic configuration is perfectly permissible. In this special case the double integral in \eqref{A5} factorizes and equals
\begin{align}
 &A_5^{(v)}\Big|_{s_{23}=0}
 \hspace{-2mm}
=
\Gamma_v(1+s_{12})
\Gamma_v(1+s_{24})
\Gamma_v(-1 \hspace{-0.5mm}-\hspace{-0.5mm}s_{12}\hspace{-0.5mm}-\hspace{-0.5mm}s_{24})\,
\Gamma_v(1+s_{13})
\Gamma_v(1+s_{34})
\Gamma_v(-1\hspace{-0.5mm}- \hspace{-0.5mm}s_{13}\hspace{-0.5mm}-\hspace{-0.5mm}s_{34})
\nonumber
\\[12pt]
\label{A5special}
&=
\frac{\zeta_v(1+s_{12})\zeta_v(1+s_{24})\zeta_v(-1-s_{12}-s_{24})}{\zeta_v(-s_{12})\zeta_v(-s_{24})\zeta_v(2+s_{12}+s_{24})}
\,
\frac{\zeta_v(1+s_{13})\zeta_v(1+s_{34})\zeta_v(-1-s_{13}-s_{34})}{\zeta_v(-s_{13})\zeta_v(-s_{34})\zeta_v(2+s_{13}+s_{34})}\,.
\end{align}
Since $A_5^{(v)}\big|_{s_{23}=0}$ is given by a product of local gamma functions, it is tempting to conclude that $A_5^{(\infty)}\prod_{p\in\mathbb{P}}A_5^{(p)}\big|_{s_{23}=0}$ is equal to one, but it can be numerically checked that this is not the case. In fact the product can be evaluated analytically through the use of a simple algebraic identity. From the definition \eqref{zetap} of the local zeta function, it follows that for any $z,w\in\mathbb{C}$:
\begin{align}
\frac{\zeta_p(z)\zeta_p(w)}{\zeta_p(z+w)}
=\frac{1-p^{-z-w}}{(1-p^{-z})(1-p^{-w})}
=-\frac{1-p^{z+w}}{(1-p^z)(1-p^w)}
=
-\frac{\zeta_p(-z)\zeta_p(-w)}{\zeta_p(-z-w)}\,.
\label{simpleid}
\end{align}
By twice applying this identity to $A_5^{(v)}\big|_{s_{23}=0}$, we find that 
\begin{align}
&\hspace{65mm}A_5^{(v)}\Big|_{s_{23}=0}
=\,
\label{A5I}
\\[6pt]
&
\frac{\zeta_v(-1-s_{12})\zeta_v(-1-s_{24})\zeta_v(-1-s_{12}-s_{24})}{\zeta_v(-s_{12})\zeta_v(-s_{24})\zeta_v(-2-s_{12}-s_{24})}
\,
\frac{\zeta_v(-1-s_{13})\zeta_v(-1-s_{34})\zeta_v(-1-s_{13}-s_{34})}{\zeta_v(-s_{13})\zeta_v(-s_{34})\zeta_v(-2-s_{13}-s_{34})}\,. \nonumber
\end{align}
Suppose now we are in the following kinematic regime:
\begin{align}
\text{(I)}&\hspace{10mm} s_{23}=0\,,\hspace{10mm}
s_{12},s_{24},s_{13},s_{34}<-2\,.
\label{I}
\end{align}
The advantage to so doing is that the convergence requirement Re$[z]>1$ is satisfied for each local zeta function $\zeta_p(z)$ in \eqref{A5I}, so that we can reliably infer that 
\begin{align}
&\hspace{62mm}
\prod_{p\in\mathbb{P}}A_5^{(v)}\Big|_{\text{(I)}}
=\,
\\
&
\frac{\zeta(-1-s_{12})\zeta(-1-s_{24})\zeta(-1-s_{12}-s_{24})}{\zeta(-s_{12})\zeta(-s_{24})\zeta(-2-s_{12}-s_{24})}
\,
\frac{\zeta(-1-s_{13})\zeta(-1-s_{34})\zeta(-1-s_{13}-s_{34})}{\zeta(-s_{13})\zeta(-s_{34})\zeta(-2-s_{13}-s_{34})}\,.
\nonumber
\end{align}
By multiplying this expression with the real 5-point amplitude, we find that the adelic 5-point amplitude in region (I) is given by
\begin{align}
A_5^{(\mathbb{A})}\Big|_{\text{(I)}}=\,&A_5^{(\infty)}\prod_{p\in\mathbb{P}}A_5^{(v)}\Big|_{\text{(I)}}
\label{A5Iadelic}
\\
=\,&
\frac{\zeta(-1-s_{12})\zeta(-1-s_{24})\zeta(2+s_{12}+s_{24})}{\zeta(1+s_{12})\zeta(1+s_{24})\zeta(-2-s_{12}-s_{24})}
\,
\frac{\zeta(-1-s_{13})\zeta(-1-s_{34})\zeta(2+s_{13}+s_{34})}{\zeta(1+s_{13})\zeta(1+s_{34})\zeta(-2-s_{13}-s_{34})}\,.
\nonumber
\end{align}
Region (I) is not the only regime in which the adelic amplitude converges and admits a closed-form answer. The same arguments used above can be applied in the eight following instances to produce the results listed below:
\begin{align}
\text{(II)}&\hspace{10mm} s_{23}=0\,,\hspace{10mm}
s_{13}\,,\,s_{24}\,,\,s_{34}<-2\,,
\hspace{10mm}
s_{12}>-s_{24}\,,
\\[6pt]
A_5^{(\mathbb{A})}\Big|_{\text{(II)}}
&=
\frac{\zeta(-s_{12})\zeta(-1-s_{24})\zeta(1+s_{12}+s_{24})}{\zeta(s_{12})\zeta(1+s_{24})\zeta(-1-s_{12}-s_{24})}\,
\frac{\zeta(-1-s_{13})\zeta(-1-s_{34})\zeta(2+s_{13}+s_{34})}{\zeta(1+s_{13})\zeta(1+s_{34})\zeta(-2-s_{13}-s_{34})}\,,
\nonumber
\end{align}
\begin{align}
\text{(III)}&\hspace{10mm} s_{23}=0\,,\hspace{10mm}
s_{12}\,,\,s_{13}\,,\,s_{34}<-2\,,
\hspace{10mm}
s_{24}>-s_{12}\,,
\\[6pt]
A_5^{(\mathbb{A})}\Big|_{\text{(III)}}
&=
\frac{\zeta(-1-s_{12})\zeta(-s_{24})\zeta(1+s_{12}+s_{24})}{\zeta(1+s_{12})\zeta(s_{24})\zeta(-1-s_{12}-s_{24})}\,
\frac{\zeta(-1-s_{13})\zeta(-1-s_{34})\zeta(2+s_{13}+s_{34})}{\zeta(1+s_{13})\zeta(1+s_{34})\zeta(-2-s_{13}-s_{34})}\,,
\nonumber
\end{align}
\begin{align}
\text{(IV)}&\hspace{10mm} s_{23}=0\,,\hspace{10mm}
s_{12}\,,\,s_{24}\,,\,s_{34}<-2\,,
\hspace{10mm}
s_{13}>-s_{34}\,,
\\[6pt]
A_5^{(\mathbb{A})}\Big|_{\text{(IV)}}
&=
\frac{\zeta(-1-s_{12})\zeta(-1-s_{24})\zeta(2+s_{12}+s_{24})}{\zeta(1+s_{12})\zeta(1+s_{24})\zeta(-2-s_{12}-s_{24})}\,
\frac{\zeta(-s_{13})\zeta(-1-s_{34})\zeta(1+s_{13}+s_{34})}{\zeta(s_{13})\zeta(1+s_{34})\zeta(-1-s_{13}-s_{34})}\,,
\nonumber
\end{align}
\begin{align}
\text{(V)}&\hspace{10mm} s_{23}=0\,,\hspace{10mm}
s_{12}\,,\,s_{13}\,,\,s_{24}<-2\,,
\hspace{10mm}
s_{34}>-s_{13}\,,
\\[6pt]
A_5^{(\mathbb{A})}\Big|_{\text{(V)}}
&=
\frac{\zeta(-1-s_{12})\zeta(-1-s_{24})\zeta(2+s_{12}+s_{24})}{\zeta(1+s_{12})\zeta(1+s_{24})\zeta(-2-s_{12}-s_{24})}\,
\frac{\zeta(-1-s_{13})\zeta(-s_{34})\zeta(1+s_{13}+s_{34})}{\zeta(1+s_{13})\zeta(s_{34})\zeta(-1-s_{13}-s_{34})}\,,
\nonumber
\end{align}
\begin{align}
\text{(VI)}&\hspace{10mm} s_{23}=0\,,\hspace{10mm}
s_{24}\,,\,s_{34}<-2\,,
\hspace{10mm}
s_{12}>-s_{24}\,,
\hspace{10mm}
s_{13}>-s_{34}\,,
\\[6pt]
A_5^{(\mathbb{A})}\Big|_{\text{(VI)}}
&=
\frac{\zeta(-s_{12})\zeta(-1-s_{24})\zeta(1+s_{12}+s_{24})}{\zeta(s_{12})\zeta(1+s_{24})\zeta(-1-s_{12}-s_{24})}\,
\frac{\zeta(-s_{13})\zeta(-1-s_{34})\zeta(1+s_{13}+s_{34})}{\zeta(s_{13})\zeta(1+s_{34})\zeta(-1-s_{13}-s_{34})}\,,
\nonumber
\end{align}
\begin{align}
\text{(VII)}&\hspace{10mm} s_{23}=0\,,\hspace{10mm}
s_{13}\,,\,s_{24}<-2\,,
\hspace{10mm}
s_{12}>-s_{24}\,,
\hspace{10mm}
s_{34}>-s_{13}\,,
\\[6pt]
A_5^{(\mathbb{A})}\Big|_{\text{(VII)}}
&=
\frac{\zeta(-s_{12})\zeta(-1-s_{24})\zeta(1+s_{12}+s_{24})}{\zeta(s_{12})\zeta(1+s_{24})\zeta(-1-s_{12}-s_{24})}\,
\frac{\zeta(-1-s_{13})\zeta(-s_{34})\zeta(1+s_{13}+s_{34})}{\zeta(1+s_{13})\zeta(s_{34})\zeta(-1-s_{13}-s_{34})}\,,
\nonumber
\end{align}
\begin{align}
\text{(VIII)}&\hspace{10mm} s_{23}=0\,,\hspace{10mm}
s_{12}\,,\,s_{34}<-2\,,
\hspace{10mm}
s_{24}>-s_{12}\,,
\hspace{10mm}
s_{13}>-s_{34}\,,
\\[6pt]
A_5^{(\mathbb{A})}\Big|_{\text{(VIII)}}
&=
\frac{\zeta(-1-s_{12})\zeta(-s_{24})\zeta(1+s_{12}+s_{24})}{\zeta(1+s_{12})\zeta(s_{24})\zeta(-1-s_{12}-s_{24})}\,
\frac{\zeta(-s_{13})\zeta(-1-s_{34})\zeta(1+s_{13}+s_{34})}{\zeta(s_{13})\zeta(1+s_{34})\zeta(-1-s_{13}-s_{34})}\,,
\nonumber
\end{align}
\begin{align}
\text{(IX)}&\hspace{10mm} s_{23}=0\,,\hspace{10mm}
s_{12}\,,\,s_{13}<-2\,,
\hspace{10mm}
s_{24}>-s_{12}\,,
\hspace{10mm}
s_{34}>-s_{13}\,,
\label{A5IXadelic}
\\[6pt]
A_5^{(\mathbb{A})}\Big|_{\text{(IX)}}
&=
\frac{\zeta(-1-s_{12})\zeta(-s_{24})\zeta(1+s_{12}+s_{24})}{\zeta(1+s_{12})\zeta(s_{24})\zeta(-1-s_{12}-s_{24})}\,
\frac{\zeta(-1-s_{13})\zeta(-s_{34})\zeta(1+s_{13}+s_{34})}{\zeta(1+s_{13})\zeta(s_{34})\zeta(-1-s_{13}-s_{34})}\,. 
\nonumber
\end{align}
It should be noted that none of the regimes (I) to (IX) are physical, in the sense that no real momenta $k_1$ to $k_5$ that are conserved, $\sum_{i=1}^5k_i=0$, and on-shell, $k_i^2=1$, will lead to kinematic invariants satisfying the conditions for any of these regimes. Take for example region (I): the inequalities $s_{12},s_{24},s_{13},s_{34}<-2$ imply that momenta $k_1$, $k_2$, $k_3$, and $k_4$ have the same sign for their zero components (all in-going or all out-going), which entails that $s_{14}<1$. But from the fact that $1=k_5^2=(\sum_{i=1}^4k_i)^2$, it follows that
\begin{align}
s_{14}=-3-s_{12}-s_{13}-s_{23}-s_{24}-s_{34}\,, \label{s14}
\end{align}
and this equation together with the conditions \eqref{I} imply that $s_{14}>5$. It is not hard to likewise show that regions (II) to (IX) are not physical either, and neither is the convergent regime studied in Ref. \cite{brekke1988non}. For the purpose of establishing that $A_5^{(\mathbb{A})}$ fails to satisfy extended analyticity, the unphysicality of the convergent kinematic configurations of course poses no issue.

\subsection{Lessons from the 5-point amplitude}
\label{2.1}

From the exact expressions \eqref{A5Iadelic} to \eqref{A5IXadelic} we are able to draw two conclusions about adelic amplitudes:

\begin{adjustwidth}{20pt}{20pt}
1) The 5-point tree-amplitude is not given by a single analytic function of the kinematic invariants $s_{ij}$. 
\end{adjustwidth}
Each of the nine expressions \eqref{A5Iadelic} to \eqref{A5IXadelic} for the adelic amplitude admits its own analytical continuation to the remaining eight regimes, but none of these analytic functions match. Analyticity breaks down even if we vary just a single kinematic invariant $s_{ij}$ and even if we restrict this invariant to real values. We conclude that the adelic amplitude is not an analytic function of the Mandelstam variables, although it is piecewise analytic in the regimes where we have been able to compute it. It is not an uncommon occurrence in number theory that functions fail to admit an analytic continuation throughout the complex plane. Consider, for example the sum $P(s)=\sum_{p\in \mathbb{P}}\frac{1}{p^s}$. For Re$[s]>1$, the sum converges absolutely, and $P(s)$ can be analytically continued into the strip with $0<\text{Re}[s]<1$. But $P(s)$ does not admit an analytic continuation to values of $s$ with non-positive real part \cite{landau1920nichtfortsetzbarkeit,froberg1968prime}; essentially, the sum ceases to be smooth as $s$ approaches the imaginary axis due to a clustering of singular points. A similar phenomenon occurs for the Dedekind eta function $\eta(\tau)=e^{\frac{\pi i \tau}{12}}\prod_{n=1}^\infty(1-e^{2n\pi i \tau})$, which cannot be analytically continued beyond the upper half-plane. The failure of analyticity of the adelic amplitude is of a different kind in that the right-hand sides of \eqref{A5Iadelic} to \eqref{A5IXadelic} are all meromorphic functions, which however only equal the adelic amplitude for restricted domains of kinematic invariants. Incidentally, Ref. \cite{frampton1988adelic} cautioned against $A_5^{(\mathbb{A})}$ possibly admitting multiple analytic continuations depending on the values of the arguments already in 1988, although this reference paradoxically made this point in order to argue that $A_5^{(\mathbb{A})}$ equals one. 
\begin{adjustwidth}{20pt}{20pt}
2) In the context of adelic amplitudes, it is not generally valid to distribute an infinite product into multiple subproducts, as in \eqref{distribute}.
\end{adjustwidth}
An identification of the form $\prod_n\big(a_n \,b_n\big) \leftrightarrow \big(\prod_n a_n\big)\big(\prod_n b_n\big)$ is only guaranteed to be valid when the subproducts $\big(\prod_n a_n\big)$ and $\big(\prod_n b_n\big)$ each converges on its own. To regulate an infinite product by splitting it into multiple products, several of which require analytic continuation, or by any other regularization method that effectuates the same splitting, is a procedure that demands careful justification, since even in some convergent cases, the value of a product can be changed by reorganizing factors. A convergent product $\prod_n(1+c_n)$ for which  $\prod_n(1+|c_n|)$ is not convergent is said to be conditionally convergent, and by a multiplicative analog of Riemann's rearrangement theorem, such products can be made to converge to any desired value by changing the order of multiplication. 

The failure of otherwise divergent, regulated products to factorize is a well-studied phenomenon in the context of operator determinants, which goes by the name of the multiplicative anomaly \cite{wodzicki1987non,kassel1989residu,kontsevich1993functional,elizalde1998zeta}. Given two operators $A$ and $B$ with eigenvalues $a_i$ and $b_i$, this anomaly is present whenever
\begin{align}
\det AB
=\prod_i a_i b_i
\neq 
\prod_i a_i\, \prod_j b_j
=
\det A\,
\det B\,,
\end{align}
where the products are understood to have been regulated. If we use zeta function regularization, we introduce a zeta function $\zeta_A(s)$ defined for sufficiently large values of Re$[s]$ by the formula
\begin{align}
\zeta_A(s)=\sum_i \frac{1}{(a_i)^s}\,,
\end{align}
and defined via analytic continuation everywhere else. Using this function, the zeta regulated determinant is given by
\begin{align}
\det A= \exp \Big[-\zeta_A'(0)\Big]\,.
\end{align}
The existence of the multiplicative anomaly amounts to the observation that there exist operators $A$ and $B$ for which
\begin{align}
\zeta'_{AB}(0)
\neq 
\zeta'_A(0)+\zeta'_B(0)\,.
\end{align}
In general, then, it is necessary to exercise caution in splitting up an infinite product. In the specific case of the adelic 5-point amplitude we see explicitly that naive application of the prescription \eqref{distribute} to the right-hand side of \eqref{A5special} results in an incorrect answer of one. Moreover, a regularization procedure that consists in separately taking an infinite product over individual factors of the local zeta function $\zeta_p(x)$ does not produce a unique result, for as the equality between \eqref{A5special} and \eqref{A5I} illustrates, there are multiple ways of expressing the $p$-adic 5-point amplitude in terms of local zeta functions.

\section{Adelic Scalar 4-Point Amplitudes}
\label{3}
In light of the lessons learned from the adelic 5-point function, it may be worthwhile to consider new ways of regulating adelic products. This section offers one such alternative and applies it to the 4-point tachyon amplitude.

\subsection{Regularization through coefficients}
\label{3.1}
Equations \eqref{A4} and \eqref{A4padic} encapsulate the dependency of the $p$-adic 4-point amplitude on the Mandelstam invariants but do not accurately fix the overall normalization, which depends on the string theory coupling constant. To allow for a different normalization than in \eqref{A4} and \eqref{A4padic}, we can dress the amplitudes $A_4^{(p)}$ with coefficients $C_p$, which may depend on $p$ but not on the Mandelstam invariants. In the presence of such coefficients, we can redefine the adelic product as
\begin{align}
A_4^{(\mathbb{A})}(s,t) =    A_4^{(\infty)}(s,t)\prod_{p\in\mathbb{P}} C_p\, A_4^{(p)}(s,t)\,.
\label{newA4}
\end{align}
This equation suggests a regularization procedure wherein, when possible, the coefficients $C_p$ are precisely chosen in such a way as to render the product convergent.\footnote{Divergent products occur in the physics literature also outside the context of adelic amplitudes. In appendix \ref{A}, it is shown how to validly apply the regularization procedure described here to a simple such example.} Such a choice of $C_p$ is not guaranteed to exist, and when it does exist, it will not be unique; we can always include additional factors $f_p$, substituting $C_p \rightarrow C_p f_p$, provided that the product $\prod_pf_p$ converges. But this ambiguity will at most contribute a momentum-independent prefactor to $A_4^{(\mathbb{A})}(s,t)$ and so does not pose a problem if we content ourselves with not determining the overall normalization of $A_4^{(\mathbb{A})}(s,t)$. A separate ambiguity, however, is introduced by the fact that different kinematic regimes require different choices of $C_p$ in order to converge, and that in some regimes no choice of coefficients results in convergence; for these latter regions it is necessary to perform a continuation from the convergent regions, which can be done in multiple ways. The upshot is that we will consider choices of values for $C_p$ that converge for different values of $s$ and $t$ as being associated to distinct candidate theories, whose merits and flaws need to be assessed separately.

\begin{figure}[t]
\begin{center}
\includegraphics[width=0.4\linewidth]{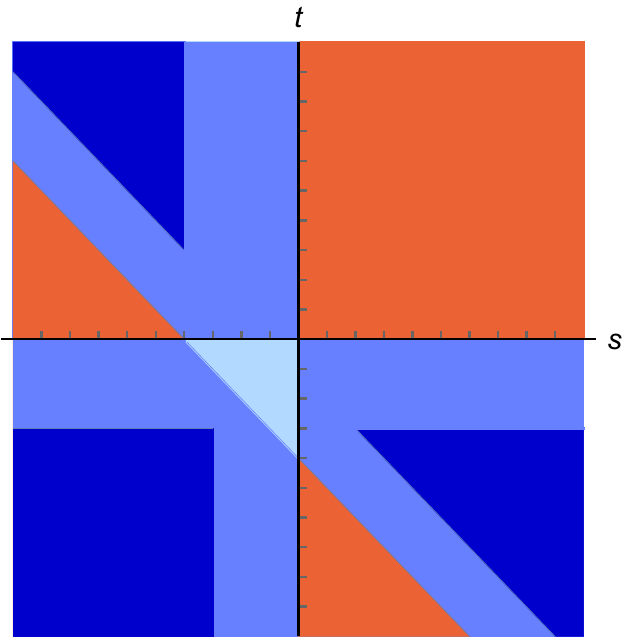}
\end{center}
\caption{Different regions of kinematic space for 4-particle tachyon scattering. The regions in {\bf \textcolor{colour2}{blue}}, {\bf \textcolor{colour1}{dark blue}}, and {\bf \textcolor{colour3}{light blue}} represent physical regions, while the {\bf\textcolor{colour4}{orange}} regions are unphysical. In the {\bf \textcolor{colour4}{orange}} regions, the product for the adelic amplitude can be rendered convergent by dressing the $p$-adic amplitudes with a coefficient of $C_p=-1$. In the {\bf \textcolor{colour1}{dark blue}} regions, convergence is achieved with coefficients $C_p=-p$. The region in {\bf \textcolor{colour3}{light blue}} marks the overlap between the $s$-, $t$-, and $u$-channels.
}
\label{fig:Regions}
\end{figure}

The string theory tachyon has a mass of $m^2=-\frac{1}{\alpha'}$, which means the Mandelstam invariants \eqref{Mandelstam} are related by
\begin{align}
s+t+u=-4\,.
\end{align}
The physical ranges of values for the Mandelstam variables of tachyonic two-to-two scattering are given as follows:
\begin{align}
\nonumber
&    \text{$s$-channel:} \hspace{3mm} s\geq -4\,, \hspace{5mm} -4-s\leq t \leq 0\,,
    \\[3pt]
&    \text{$t$-channel:} \hspace{3.5mm} t\geq -4\,, \hspace{5.5mm} -4-t\leq s \leq 0\,, \label{subsetRegions}
    \\[3pt]
&    \text{$u$-channel:} \hspace{3mm} s,t<0\,.
\nonumber
\end{align}
The union of these regions is marked in different hues of blue in figure \ref{fig:Regions}. Allowing $s$ and $t$ to assume more general values, it turns out there are two possible choices of $C_p$ that will render the product $\prod_pC_pA^{(p)}(s,t)$ convergent. For $C_p=-p$ the product is convergent in the following domains, marked in \textcolor{colour1}{\bf{dark blue}} in figure \ref{fig:Regions}:
\begin{align*}
&    \text{$s$-channel subset:} \hspace{3mm} t < -3\,, \hspace{5mm} s> - 1 - t\,,
    \\[3pt]
&    \text{$t$-channel subset:} \hspace{3mm} s < -3\,, \hspace{5mm} t> - 1 - s\,,
    \\[3pt]
&    \text{$u$-channel subset:} \hspace{3mm} s,t<-3\,.
\end{align*}
And for $C_p=-1$ the product is convergent precisely in the complement of the physical region, ie. in the following three regions, marked in \textcolor{colour4}{\bf{orange}} in figure \ref{fig:Regions}:
\begin{align}
\nonumber
&    A) \hspace{3mm} t > 0\,, \hspace{5mm} s< - 4 - t\,,
    \\[3pt]
&    B) \hspace{3mm} s > 0\,, \hspace{5mm} t< - 4 - s\,, \label{abcRegions}
    \\[3pt]
&    C) \hspace{3mm} s,t>0\,.
\nonumber
\end{align}
$\hspace{-1.5mm}$In each convergent regime, the adelic product \eqref{newA4} admits closed-form evaluation by a computation similiar to the 5-point calculation in the last section. 
For example, let us consider the case when $C_p=-p$ and $s,t<-3$. From the simple algebraic fact that
\begin{align}
-p\frac{\zeta_p(3+s+t)}{\zeta_p(2+s)\,\zeta_p(2+t)}
=
\frac{\zeta_p(-3-s-t)}{\zeta_p(-2-s)\,\zeta_p(-2-t)}\,,
\end{align}
it follows that when multiplied by coefficients $C_p=-p$, the $p$-adic amplitude can be rewritten as
\begin{align}
-pA_4^{(p)}(s,t)=
\frac{\zeta_p(-1-s)\,\zeta_p(-1-t)\,\zeta_p(-3-s-t)}{\zeta_p(-2-s)\,\zeta_p(-2-t)\,\zeta_p(-2-s-t)}\,.
\label{-pA4}
\end{align}
In the $u$-channel subset with $s,t<-3$, the product over primes converges separately for each of the six local zeta functions in \eqref{-pA4} and can consequently be evaluated straightforwardly:
\begin{align}
\prod_{p\in \mathbb{P}}\Big(-pA_4^{(p)}(s,t)\Big)=
\frac{\zeta(-1-s)\,\zeta(-1-t)\,\zeta(-3-s-t)}{\zeta(-2-s)\,\zeta(-2-t)\,\zeta(-2-s-t)}
\,.
\end{align}
Multiplying this product with the real Veneziano amplitude \eqref{VenezianoRiemann}, we arrive at the following expression for a putative non-constant $u$-channel adelic amplitude:
\begin{align}
A_4^{(\mathbb{A})}\Big|_{\text{$u$-channel}}
= A_4^{(\infty)}
\prod_{p\in \mathbb{P}}\Big(
-p A_4^{(p)}
\Big)\Big|_{\text{$u$-channel}}
=
\frac{\zeta(2+s)\,\zeta(2+t)\,\zeta(-3-s-t)}{\zeta(-2-s)\,\zeta(-2-t)\,\zeta(3+s+t)}\,.
\end{align}
Since it is not possible to attain convergence of the adelic product throughout the physical regime, we must invoke a partial analytic continuation, but a continuation of the full product, not of separate sub-products. Just as in the case of the 5-point amplitude, the result we thereby arrive at is only piecewise analytic. A peculiarity of tachyon scattering is that the $s$-, $t$-, and $u$-channels overlap. In the region of overlap, displayed in \textcolor{colour3}{\bf{light blue}} in figure \ref{fig:Regions}, the answer for $A_4^{(\mathbb{A})}$ will depend on the scattering channel. Phrased differently, $A_4^{(\mathbb{A})}$ depends not only on $s$ and $t$ but also on the signs of the time components of the momenta. This peculiarity disappears once we turn to massless scattering in the next section.

For $C_p=-p$, we continue the answers obtained in each subset of the three scattering channels to the full respective channel. For $C_p=-1$ we continue region $A)$ to the $s$-channel, $B)$ to the $t$-channel, and $C)$ to the $u$-channel. By this procedure, we arrive at the following candidate amplitudes: 
\begin{align}
C_p=-p:
\hspace{5mm}
A^{(\mathbb{A})}_4=
\begin{cases}
\displaystyle \frac{\zeta(1+s)\,\zeta(2+t)\,\zeta(2+u)}{\zeta(-1-s)\,\zeta(-2-t)\,\zeta(-2-u)}
&\hspace{5mm}\text{ $s$-channel}\,,
\\
\\
\displaystyle \frac{\zeta(2+s)\,\zeta(1+t)\,\zeta(2+u)}{\zeta(-2-s)\,\zeta(-1-t)\,\zeta(-2-u)}
&\hspace{5mm}\text{ $t$-channel}\,,
\\
\\
\displaystyle \frac{\zeta(2+s)\,\zeta(2+t)\,\zeta(1+u)}{\zeta(-2-s)\,\zeta(-2-t)\,\zeta(-1-u)}
&\hspace{5mm}\text{ $u$-channel}\,,
\end{cases}
\label{newAdelic1}
\end{align}
\begin{align}
C_p=-1:
\hspace{5mm}
A^{(\mathbb{A})}_4=
\begin{cases}
\displaystyle \frac{\zeta(2+s)\,\zeta(1+t)\,\zeta(1+u)}{\zeta(-2-s)\,\zeta(-1-t)\,\zeta(-1-u)}
&\hspace{5mm}\text{ $s$-channel}\,,
\\
\\
\displaystyle \frac{\zeta(1+s)\,\zeta(2+t)\,\zeta(1+u)}{\zeta(-1-s)\,\zeta(-2-t)\,\zeta(-1-u)}
&\hspace{5mm}\text{ $t$-channel}\,,
\\
\\
\displaystyle \frac{\zeta(1+s)\,\zeta(1+t)\,\zeta(2+u)}{\zeta(-1-s)\,\zeta(-1-t)\,\zeta(-2-u)}
&\hspace{5mm}\text{ $u$-channel}\,.
\end{cases}
\label{newAdelic2}
\end{align}
The $C_p=-1$ adelic amplitude given in \eqref{newAdelic2} is the non-constant adelic amplitude that was previously proposed in \cite{aref1988adelic}.\footnote{There is a different way to regulate the adelic product in the regions \eqref{abcRegions} without introducing a coefficient $C_p=-1$ by hand that also leads to \eqref{newAdelic2}. In the regions $A$, $B$, and $C$, the product over the $p$-adic amplitudes \eqref{A4padic} converges in modulus but the overall sign alternates between plus and minus. This means the limit set at fixed $s$ and $t$ consists of two points. If, in analogy with Ces\'aro summation, we take the geometric mean of the limit set, we recover \eqref{newAdelic2} up to an overall phase.}

Piecewise analyticity is not a customary property of tree-amplitudes, and so a natural question to ask is if there are compelling reasons why we should think of either of these expressions as representing a scattering amplitude. To address this question, we will study in turn the partial wave decompositions and high energy asymptotics of \eqref{newAdelic1} and \eqref{newAdelic2}.

\subsection{Partial wave decomposition}
\label{3.2}
Scattering amplitudes in quantum theories admit of partial wave decompositions into Gegenbauer polynomials $C_m^{(\alpha)}(x)$. Viewed abstractly, the Gegenbauer polynomials form an infinite family of one-parameter polynomials satisfying the orthogonality relation
\begin{align}
\int_{-1}^1dx\,C_m^{(\alpha)}(x)\,C_n^{(\alpha)}(x)\,(1-x^2)^{\alpha-\frac{1}{2}}
=\delta_{m,n}\frac{2^{1-2\alpha}\pi\Gamma(2\alpha+n)}{n!(n+\alpha)\Gamma(\alpha)^2}\,.
\label{GegOrtho}
\end{align}
The index $m$ takes values in $\mathbb{N}_0$ and labels the degree of a given polynomial, while $\alpha$ is a real parameter that in the context of partial wave decompositions assumes the value $\alpha = (d-3)/2$. In four dimensions the Gegenbauer polynomials  reduce to the Legendre polynomials $C_m^{(\frac{1}{2})}(x)$, and in five dimensions they reduce to the Chebyshev polynomials $C_m^{(1)}(x)$. For even $m$, $C_m^{(\alpha)}(x)$ contains only even powers of $x$, for odd $m$, only odd. The first four Gegenbauer polynomials are given by
\begin{align}
C_0^{(\alpha)}(x)=&\,1\,,  \\
C_1^{(\alpha)}(x)=&\,2\alpha\, x\,, \\
C_2^{(\alpha)}(x)=&\,2\alpha(1+\alpha)x^2-\alpha\,, \\
C_3^{(\alpha)}(x)=&\,\frac{4}{3}\alpha(1+\alpha)(2+\alpha)x^3-2\alpha(1+\alpha)x\,.
\end{align}
Unitarity dictates that any residue of a four-point amplitude $A_4(s,t)$ decomposes into a positively-weighted sum of Gegenbauer polynomials $C_m^{(\frac{d-3}{2})}(\cos\theta)$, where $\theta$ is the scattering angle in the center-of-mass frame. Such a frame does not always exist for tachyonic scattering, but we can still perform the decomposition
\begin{align}
\underset{s=s^\ast}{\text{Res}}\,A_4\left(s,\frac{(s+4)(\cos\theta-1)}{2}\right)
=\sum_{m=\mathbb{N}_0}K^{(s^\ast)}_m(d)\,C_m^{(\frac{d-3}{2})}(\cos\theta)\,,
\label{Geg}
\end{align}
where all non-zero coefficients $K^{(s^\ast)}_m(d)$ must be positive for unitarity to hold.\footnote{We refer the reader for to the appendices of Caron-Huot, Komargodski, Sever, and Zhiboedov's paper \cite{caron2017strings} for a nice review of partial wave unitarity with application to the Veneziano amplitude.}

Let us review how this decomposition works for the real Veneziano amplitude $A_4^{(\infty)}(s,t)$, which has $s$-channel poles at $s=s^\ast \in 2\mathbb{N}_0-1$, and for which the number of non-zero Gegenbauer coefficients in generic number of spacetime dimensions grows linearly as $(s^\ast+3)/2$. There are no poles at even values of $s$, since open string excitations of even mass in units of $\frac{1}{\alpha'}$ carry odd spin, and we are studying the full rather than the partial amplitude. In order for the coefficients $K_m^{(s^\ast)}$ in \eqref{Geg} to be positive, we must equip $A_4^{(\infty)}(s,t)$ with an overall minus sign.\footnote{To make positive the residues of the $p$-adic amplitude $A_4^{(p)}(s,t)$, it must also be multiplied with a negative prefactor. Conceivably this fact helps account for the negativity of the coefficients $C_p$. 
}
The first residue trivially decomposes into a Gegenbauer polynomial:
\begin{align}
&\underset{s=-1}{\text{Res}}\left[-A_4^{(\infty)}\left(s,\frac{(s+4)(\cos\theta-1)}{2}\right)\right]=\,
2=2\,C_0^{(\frac{d-3}{2})}(\cos\theta)\,.
\end{align}
The second residue decomposes as
\begin{align}
\underset{s=1}{\text{Res}}\left[-A_4^{(\infty)}\left(s,\frac{(s+4)(\cos\theta-1)}{2}\right)\right]
=\,&\frac{1}{4}(25\cos^2\theta-1)
\\[6pt]
=\,&
\frac{26-d}{4(d-1)}\,
C_0^{(\frac{d-3}{2})}(\cos\theta)
+
\frac{25}{2d^2-8d+6}\,
C_2^{(\frac{d-3}{2})}(\cos\theta)\,.
\nonumber
\end{align}
From the coefficient $K_0^{(1)}(d)$, we observe that the Veneziano amplitude violates unitarity for $d>26$ as first observed by Frampton \cite{frampton1972n} in 1972, in accordance with Lovelace's discovery of the previous year that 26 is the critical dimension of bosonic string theory \cite{lovelace1971pomeron}. The no-ghost theorem implies the positivity of all coefficients for $d\leq 26$, but no direct proof is known.\footnote{Ref. \cite{maity2022positivity} offered a partial proof for the case $d=4$.} However, Arkani-Hamed, Eberhardt, Huang, and Mizera \cite{arkani2022unitarity} recently developed technology that enabled them to provide such a direct proof when $d\leq 10$ and also to directly prove the positivity of the Gegenbauer coefficients for all superstring amplitudes in $d\leq 6$, and similar methods were successfully applied to the $q$-deformed version of the Veneziano amplitude known as the Coon amplitude \cite{coon1969uniqueness} by Bhardwaj, De, Spradlin, and Volovich \cite{bhardwaj2022unitarity}.

For general string amplitudes, Gegenbauer coefficients do not have to become negative immediately above the critical dimension. For the Virasoro-Shapiro amplitude, numerical evaluation of the Gegenbauer coefficients for the first many residues indicates that partial wave positivity is not violated until the number of spacetime dimensions exceeds 57.

\subsubsection*{$C_p=-p$ adelic amplitude}

In equation \eqref{newAdelic1}, the $C_p=-p$ candidate 4-point adelic amplitude was written down in terms of three expressions, one for each scattering channel. In actuality this notation is redundant, and without loss of generality, we are free to take the zero components $k_1^0$ and $k_2^0$ to be positive and $k_3^0$ and $k_4^0$ to be negative, so that we are in the $s$-channel.

In the $s$-channel subset depicted in {\bf \textcolor{colour1}{dark blue}} in  figure \ref{fig:Regions}, where the $C_p=-p$ adelic product is convergent, the adelic amplitude shares its poles with those of the real Veneziano amplitude, meaning there are poles at $s\in 2\mathbb{N}+1$. Unlike the residues of the real Veneziano amplitude, the residues of $A_4^{(\mathbb{A})}$ are not polynomials in $\cos\theta$, for which reason the coefficients $K_m^{(s^\ast)}$ do not admit straightforward closed-form evaluation. Instead we resort to numeric computation via the formula
\begin{align}
K_m^{(s^\ast)}\hspace{-0.3mm}=\hspace{-0.3mm}
\frac{m!(2m+d-3)\Gamma\big(\frac{d-3}{2}\big)^2}{2^{5-d}\pi\Gamma(d-3+m)}
\hspace{-0.6mm}\int_{-1}^1 \hspace{-0.6mm}dx\,C_m^{(\frac{d-3}{2})}(x)\, \underset{s=s^\ast}{\text{Res}}\Big[\hspace{-0.5mm}-\hspace{-0.6mm}A^{(\mathbb{A})}_4\Big(s,\frac{(s+4)(x-1)}{2}\Big)\Big](1-x^2)^{\frac{d-4}{2}}
\,,
\nonumber
\end{align}
where $A_4^{(\mathbb{A})}$, whose overall normalization is undetermined, has been dressed with a prefactor of minus one, which will be needed for unitarity. Also unlike the real Veneziano amplitude, the $s$-channel residues of $A^{(\mathbb{A})}_4$ decompose into infinite rather than finite sums over Gegenbauer polynomials, an unorthodox property also present in the UV-complete gravity amplitude recently discovered by Huang and Remmen \cite{huang2022uv} and pointed out by Geiser and Lindwasser \cite{geiser2022generalized} to be present in generalizations of the Veneziano amplitude.

If we analytically extend the $C_p=-p$ amplitude to the entirety of the $s$-channel, we encounter new poles: at $s=1$, at $t,u=-1$, and at $s,t,u=0$. The tachyonic pole in $s$ is absent: only an in-going and an out-going tachyon can together generate a tachyonic resonance in this set-up. The poles at $s=0$ and $t,u=-1$ each have a constant residue of 12. The fact that the $s=0$ pole is a constant, unlike all the other $s$-channel poles, which are non-polynomial in $\cos\theta$, is a fortuitous circumstance, since the adelic amplitude thereby eschews exchanges of higher spins on the massless pole.

At the intersection of the poles at $s\in 2\mathbb{N}_0+1$ and at $t=0$ or $u=0$  we encounter problematic double poles. One consequence of these is the breakdown of the Gegenbauer decomposition for $d\leq 4$, since the numerical integral $K_m^{(s^\ast)}$ ceases to converge in low dimensions. The presence of double poles is certainly alarming, but an extenuating circumstance is provided by the fact that they are situated on the edge of the physically allowed region. It would be further problematic if we attempted to analytically continue $A_4^{(\mathbb{A})}$ into the complex plane, since the non-trivial zeros of the Riemann zeta function would give rise to complex poles.

\begin{figure}
\begin{center}
\includegraphics[width=0.99\linewidth]{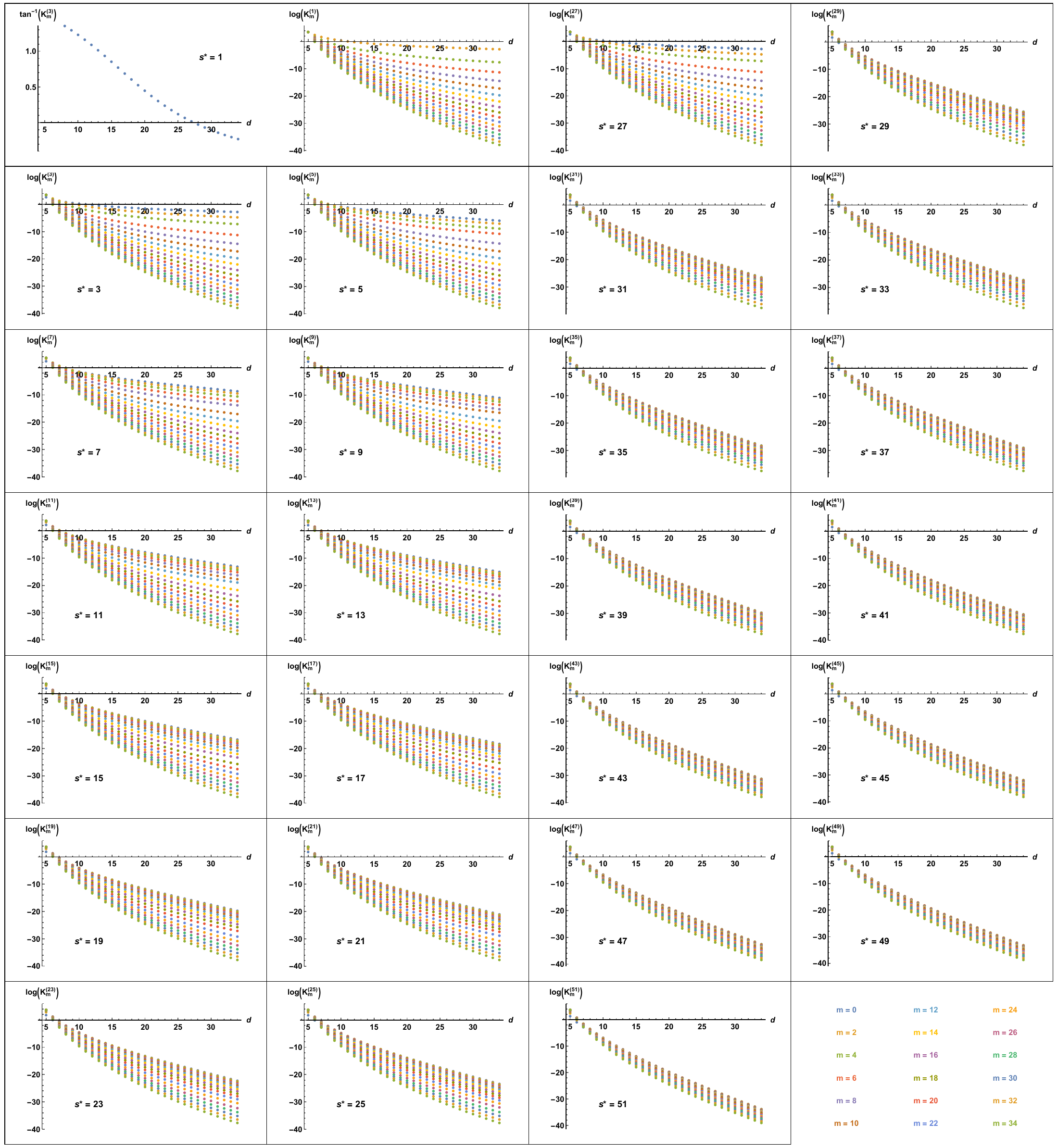}
\end{center}
\caption{Coefficients in the $s$-channel Gegenbauer decomposition of the tentative adelic amplitude with $C_p=-p$. All the coefficients are positive with the exception of $K^{(1)}_{0}(d)$, which falls negative between $d=27$ and $d=28$. For $d\leq 4$, the coefficients blow up. For the sake of visibility, the logarithm has been taken of coefficients $K^{(s^\ast)}_{m}(d)$ that are positive for all $d>4$.
}
\label{fig:coeffs1}
\end{figure}

While the above-mentioned issues cast doubt on the interpretation of $A_4^{(\mathbb{A})}$ as a scattering amplitude, there is also evidence to the contrary. There is no reason why Gegenbauer coefficients for different polynomials and at different poles should have the same sign without unitarity to enforce this constraint. Yet for $5\leq d \leq 27$, numerical evidence suggests that the coefficients are all positive. Figure \ref{fig:coeffs1} shows the values of the first 18 Gegenbauer coefficients at the first 25 $s$-channel poles. The coefficient of the zeroth Gegenbauer polynomial at the first pole is positive for $4<d\leq 27$ and falls negative between $d=27$ and $d=28$, while all other coefficients are positive throughout the range of dimensions explored.

\begin{figure}
\begin{center}
\includegraphics[width=0.99\linewidth]{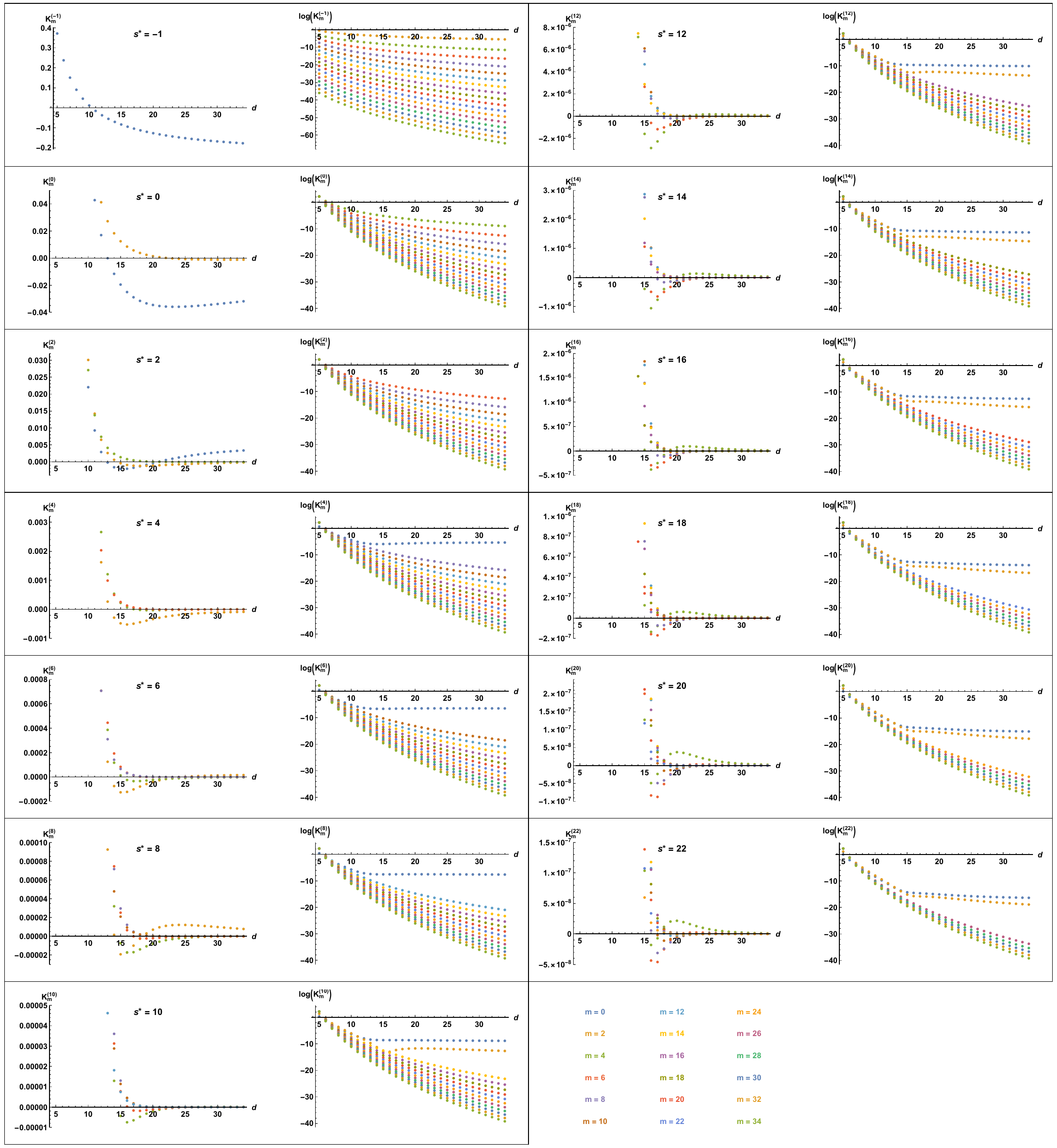}
\end{center}
\caption{Coefficients in the $s$-channel Gegenbauer decomposition of the tentative adelic amplitude with $C_p=-1$. In five to ten dimensions all the coefficients are positive. In 11 dimensions and higher, some coefficients are negative. For $d\leq 4$, the coefficients blow up. For the sake of visibility, the logarithm has been taken of coefficients $K^{(s^\ast)}_{m}(d)$ that are positive for all $d>4$.
}
\label{fig:coeffs2}
\end{figure}

\subsubsection*{$C_p=-1$ adelic amplitude}

In the $s$-channel region, the $C_p=-1$ adelic amplitude has poles at $s=-1$, for $s\in 2\mathbb{N}_0$, and at $t,u=0$. This means that in a hypothetical adelic string theory spectrum, the spin-even particles would carry even rather than odd masses in units of $\frac{1}{\alpha'}$. As with the $C_p=-p$ adelic amplitude, we encounter problematic double poles when the above equations for $s$ and $t$ or $u$ are simultaneously satisfied, and again these double poles entail that we can only perform the Gegenbauer decomposition \eqref{Geg} when $d>4$. And as before, imposing extended analyticity would result in complex poles corresponding to the non-trivial zeros of the Riemann zeta function.

The values of the first 18 Gegenbauer coefficients at the first 13 poles in $s$ are shown in figure \ref{fig:coeffs2}. Generally the coefficients $K_m^{(s^\ast)}(d)$ are not monotonic functions of $m$, $s^\ast$, or $d$. With increasing $s^\ast$, there are an increasing number of values of $m$ for which $K_m^{(s^\ast)}(d)$ assumes negative values, but in no case is $K_m^{(s^\ast)}(d)$ negative for $d\leq 10$. The first unitarity violation occurs when $K^{(-1)}_0(d)$ falls negative between $d=10$ and $d=11$. For $4<d\leq 10$ the numerically computed coefficients $K_m^{(s^\ast)}(d)$ are all positive. Unlike the case for the $C_p=-p$ amplitude, to achieve positivity, the plotted Gegenbauer coefficients are associated to $A^{(\mathbb{A})}_4$ rather than $-A^{(\mathbb{A})}_4$. Another difference from the $C_p=-p$ amplitude, and very possibly a serious malady, is the presence of an infinite tower of Gegenbauer polynomials at the massless residue.

\subsection{High energy limit}
\label{3.3}
A desirable property of the Veneziano amplitude is its well-behaved high energy asymptotics, as nicely reviewed in the first chapter of Green-Schwarz-Witten \cite{green}. For external scalars scattering and exchanging an internal particle of spin $J$ at large $s$ and fixed $t$, the tree-amplitude has the form
\begin{align}
A_J(s,t) \approx  -\frac{g^2(-s)^J}{t-M^2}\,. \label{J}
\end{align}
By unitarity, loop amplitudes can be constructed by sewing together tree-amplitudes. But for $d\geq 4$, the high energy scaling \eqref{J} gives rise to unrenormalizable divergences when $J>1$, which is problematic if we want a theory with particles of spin greater than one. String theory comes to the rescue by furnishing us with an infinite tower of particles with increasingly higher spin that resums to a more benign high energy behaviour. In the Regge limit and the fixed angle high energy limit, using the asympotic behaviour of the gamma function, 
\begin{align}
x>>1:
\hspace{15mm}
\Gamma(x)\approx \sqrt{\frac{2\pi}{x}}\,\Big(\frac{x}{e}\Big)^{x}\,,
\hspace{15mm}
\Gamma(-x)\approx
-\sqrt{\frac{\pi}{2x}}\,\Big(\frac{x}{e}\Big)^{-x}\csc(\pi x)\,,
\label{gammaAsymptotic}
\end{align}
it can be shown that the Veneziano amplitude exhibits the following asymptotics:
\begin{align}
&\text{large $s$, fixed $t$:} \hspace{12mm}
A_4^{(\infty)}(s,t) \approx
2\,s^{1+t}\,\Gamma(-1-t)
\sec\Big(\frac{\pi s}{2}\Big)
\sin\Big(\frac{\pi t}{2}\Big)
\sin\Big(\frac{\pi (s+t)}{2}\Big)\,,
\nonumber \\[8pt] \label{decay}
&\text{large $s$, fixed $\theta$:} \hspace{12mm}
A_4^{(\infty)}\Big(s,\frac{s+4}{2}(\cos\theta-1)\Big)\approx
\\[5pt]
&
\bigg|\frac{2}{\sin\theta}\tan(\frac{\theta}{2})^{\cos\theta}\bigg|^{-s}
\,
\sqrt{\frac{\pi\sin^2\theta}{2s}}
\,
\Big(\cot^2\big(\frac{\theta}{2}\big)\Big)^{2\cos\theta}
\bigg(
1-\cos\Big(\frac{\pi(4+s)\cos\theta}{2}\Big)
\sec\Big(\frac{\pi s}{2}\Big)
\bigg)\,. \nonumber
\end{align}
We see that $A_4^{(\infty)}$ scales as $s^{1+t}$ in the Regge limit and decays exponentially in the fixed angle large energy limit. 

To investigate whether the candidate adelic amplitudes \eqref{newAdelic1} and \eqref{newAdelic2} exhibit similar desirable high energy limits, it is convenient to write each adelic amplitudes as the product of the Veneziano amplitude and the piece due to the $p$-adic amplitudes: 
\begin{align}
A_4^{(\mathbb{A})}\big|_{s\text{-channel}}(s,t)=A_4^{(\infty)}(s,t)\,A_4^{(\mathbb{P})}(s,t)\big|_{s\text{-channel}}\,,
\end{align}
where the piece relating the adelic and real amplitudes is given by
\begin{align}
&
C_p=-p:
\hspace{12mm}
A^{(\mathbb{P})}_4(s,t)\big|_{s\text{-channel}}
=\frac{
\zeta(1+s)\,
\zeta(-1-t)\,
\zeta(3+s+t)
}{
\zeta(2+s)\,
\zeta(-2-t)\,
\zeta(2+s+t)
}\,,
\hspace{16mm}
\\[4pt]
&
C_p=-1:
\hspace{12mm}
A^{(\mathbb{P})}_4(s,t)\big|_{s\text{-channel}}
=\frac{
\zeta(-1-s)\,
\zeta(1+t)\,
\zeta(-3-s-t)
}{
\zeta(-2-s)\,
\zeta(2+t)\,
\zeta(-2-s-t)
}\,.
\hspace{16mm}
\end{align}
Since $\zeta(x)\rightarrow 1$ as $x \rightarrow \infty$, we immediately see that the $C_p=-p$ adelic amplitude has the same high energy asymptotics as $A_4^{(\infty)}(s,t)$. For the $C_p=-1$ adelic amplitude, it takes slightly more work to determine the asympotics. But by invoking the functional equation for the Riemann zeta function, Euler's reflection formula for the gamma function, and a bit of trigonometry, one finds that
\begin{align}
\label{pFactor2}
&
C_p=-1:
\hspace{48mm}
A^{(\mathbb{P})}_4(s,t)\big|_{s\text{-channel}}=
\\[6pt]
&\hspace{6mm}
-\frac{2(1+t)(3+s+t)}{\pi(2+s)}
\frac{
\cos\big(\frac{\pi s}{2}\big)
\cos\big(\frac{\pi t}{2}\big)
\cos\big(\pi\frac{s+t}{2}\big)
}
{
\sin\big(\pi s\big)+\sin\big(\pi t\big)-\sin\big(\pi(s+t)\big)
}
\frac{
\zeta(2+s)\,
\zeta(-t)\,
\zeta(4+s+t)
}{
\zeta(3+s)\,
\zeta(-1-t)\,
\zeta(3+s+t)
}\,.
\nonumber
\end{align}
In the limit of large $s$ and fixed $t$, this expression tends to a periodically oscillating function of $s$, and in the limit of large $s$ and fixed $\theta$, \eqref{pFactor2} tends to a periodic function of $s$ times a function that grows linearly in $s$. The linear growth is subdominant compared to the exponential decay in \eqref{decay}, and so the healthy high energy behaviour of the real amplitude persists also for the $C_p=-1$ adelic amplitude.

\subsection{Interpretation as integral over ring of adeles}
\label{3.4}
The term ``adelic" applied to product formulas and amplitudes is motivated by a tentative connection to the adelic numbers. This subsection discusses this connection and how it relates to the choices of $C_p$ introduced above.

The ring $\mathbb{A}$ of adeles over $\mathbb{Q}$ is given by the restricted product
\begin{align}
\mathbb{A}=\mathbb{R}\times \prod_{p\in\mathbb{P}}{}'\,\, \mathbb{Q}_p\,.
\end{align}
What this means concretely is that the adelic numbers are given by the following set:
\begin{align}
\nonumber
\mathbb{A}=\bigg\{
(x_\infty,\,x_2,\,x_3,\,x_5,...)\hspace{3mm}\Big|\hspace{3mm} &x_\infty \in \mathbb{R}\,,
\\[-2pt]
& x_p \in \mathbb{Q}_p \text{ for all } p\in \mathbb{P}\,,
\\
& x_p \in \mathbb{Z}_p \text{ for all but finitely many } p\in \mathbb{P}
\bigg\}\,.
\nonumber
\end{align}
An infinite product like \eqref{newA4} is referred to as an adelic amplitude because ideally it admits an interpretation as an integral over the adelic numbers:
\begin{align}
A_4^{(\mathbb{A})}(s,t)
\stackrel{?}{=}
\int_{\mathbb{A}}dx\,|x|_{\mathbb{A}}^{-2-s}|1-x|_{\mathbb{A}}^{-2-t}\,,
\label{adelicIntegral}
\end{align}
where the adelic norm is defined as follows:
\begin{align}
|x|_{\mathbb{A}}=|x_\infty|_\infty\prod_{p\in \mathbb{P}}|x_p|_p\,.
\end{align}
The condition on an adelic number that all but finitely many of the coordinates $x_v$ belong to $\mathbb{Z}_p$ indicates that we cannot immediately factor an adelic integral into an infinite product of integrals over $\mathbb{Q}_p$. A more careful way to carry out the integral would be to partition the set of all primes into two subsets, $\mathbb{P}=P_1 \cup P_2$, where $P_1$ is finite and $P_2$ is infinite, and then integrate primes in $P_1$ over $\mathbb{Q}_p$ and primes in $P_2$ over $\mathbb{Z}_p$, before finally taking the limit as $P_1$ tends to $\mathbb{P}$:
\begin{align}
A_4^{(\mathbb{A})}(s,t)\stackrel{?}{=}
\lim_{P_1\rightarrow \mathbb{P}}
\bigg[
\prod_{p\in P_1}
\int_{\mathbb{Q}_p}dx\,|x|_p^{-2-s}|1-x|_p^{-2-t}
\bigg]\,
\bigg[
\prod_{p \in P_2}
\int_{\mathbb{Z}_p}dx\,|x|_p^{-2-s}|1-x|_p^{-2-t}
\bigg]\,.
\end{align}
For such a procedure to be well-defined, the precise details of how we partition $\mathbb{P}$ into $P_1$ and $P_2$ and take the limit $P_1\rightarrow \mathbb{P}$ should not matter. For the procedure to work we must also require that the product 
\begin{align}
\prod_{p\in\mathbb{P}}\int_{\mathbb{Z}_p}dx\,|x|_p^{-2-s}|1-x|_p^{-2-t}
\end{align}
be well-defined. But we note that
\begin{align}
\int_{\mathbb{Z}_p}dx\,|x|_p^{-2-s}|1-x|_p^{-2-t}
=\frac{-2+p+p^{1+s}+p^{1+t}-p^{3+s+t}}{p(p^{1+s}-1)(p^{1+t}-1)}\,,
\end{align}
and the product of this expression over all primes tends to zero or infinity, depending on the values of $s$ and $t$. We can however render the product convergent by equipping the integrals with coefficients. In particular, in the kinematic regimes \eqref{subsetRegions} and \eqref{abcRegions}, we can use precisely the respective constants $C_p=-p$ and $C_p=-1$ to attain convergence of the product
\begin{align}
\prod_{p\in\mathbb{P}} C_p \int_{\mathbb{Z}_p}dx\,|x|_p^{-2-\alpha's}|1-x|_p^{-2-\alpha't}\,.
\end{align}
With the coefficients in place, we can identify the adelic integral as
\begin{align}
A^{(\mathbb{A})}_4(s,t)
=
\lim_{P_1\rightarrow \mathbb{P}}
\bigg[
\prod_{p\in P_1}
C_p\int_{\mathbb{Q}_p}dx\,|x|_p^{-2-s}|1-x|_p^{-2-t}
\bigg]\,
\bigg[
\prod_{p\in \mathbb{P}\setminus P_1}
C_p\int_{\mathbb{Z}_p}dx\,|x|_p^{-2-s}|1-x|_p^{-2-t}
\bigg]\,. \nonumber
\end{align}
The limiting value for this expression does not depend on how the limit $P_1\rightarrow \mathbb{P}$ is taken and matches the expressions \eqref{newAdelic1} or \eqref{newAdelic2}, provided we set $C_p=-p$ or $C_p=-1$ and restrict $s$ and $t$ to the respective region of converges for these two choices of coefficients.

\section{Adelic 4-Point Superamplitudes}
\label{4}
The full four-gluon amplitude in type-I string theory is given by 
\begin{align}
\mathcal{A}_4=
K(k_i,\zeta_i) \bigg(
\frac{\Gamma(-s)\Gamma(-t)}{\Gamma(1+u)}+
\frac{\Gamma(-s)\Gamma(-u)}{\Gamma(1+t)}+
\frac{\Gamma(-t)\Gamma(-u)}{\Gamma(1+s)}
\bigg)\,,
\label{Agluon}
\end{align}
where the overall normalization and the polarization-dependent kinematic factor have been absorbed into the prefactor $K(k_i,\zeta_i)$, whose value can be read off from equations (4.23) and (4.24) of Schwarz' paper \cite{schwarz1982superstring}. Since we are now dealing with massless scattering, the Mandelstam invariants are related by
\begin{align}
s+t+u=0\,.
\end{align}
It was observed in 1989 by Ruelle, Thiran, Verstegen, and Weyers \cite{ruelle1989adelic} that, like the Veneziano amplitude, this amplitude can be expressed in terms of Gelfand-Graev gamma functions:
\begin{align}
\mathcal{A}_4=\frac{K}{2\pi i}\,\Gamma^{(-1)}_{\infty}(-s)\,\Gamma^{(-1)}_{\infty}(-t)\,\Gamma^{(-1)}_{\infty}(-u)
\equiv \frac{K}{2\pi}\,\mathcal{A}^{(\infty)}_4\,,
\label{AsuperReal}
\end{align}
where the kind of Gelfand-Graev gamma function relevant in this case is given by
\begin{align}
\Gamma^{(-1)}_{\infty}(z)=2i(2\pi)^{-z}\sin\left(\frac{\pi z}{2}\right)\Gamma(z)\,.
\end{align}
This kind of signed gamma function exists both in a real and a $p$-adic version, with the different versions defined by
\begin{align}
\Gamma^{(-1)}_\infty(z)=\int_{\mathbb{R}} \frac{dx}{|x|}\,e^{2\pi i x}\,|x|^z\,\text{sign}(x)\,,
\hspace{15mm}
\Gamma^{(-1)}_p(z)=\int_{\mathbb{Q}_p} \frac{dx}{|x|_p}\,e^{2\pi i\{ x\}}\,|x|_p^z\,\text{sign}_{-1}(x)\,,
\label{signedGammas}
\end{align}
where $\{x\}$ denotes the fractional part of $x\in \mathbb{Q}_p$ and the $p$-adic sign function is given by
\begin{align}
\text{sign}_{-1}(x)=\begin{cases}
1 \hspace{5mm}&\text{ if } x=a^2+b^2 \text{ for some }a,b\in \mathbb{Q}_p\,,
\\
-1 &\text{ otherwise}\,.
\end{cases}
\end{align}
In the $p$-adic case, carrying out the integral over $\mathbb{Q}_p$, the signed gamma function evaluates to
\begin{align}
&\Gamma_2^{(-1)}(z)=i\,2^{2z-1} \hspace{10mm}\text{for }p=2\,, \text{ while}
\label{GammaSigned}
\\[6pt]
&
\Gamma_p^{(-1)}(z)
=\frac{\zeta_p^{(-1)}(z)}{\zeta_p^{(-1)}(1-z)}\,,
\hspace{10mm}
\text{where}
\hspace{10mm}
\zeta_p^{(-1)}(z)
=\frac{1}{1+(-1)^{\frac{p+1}{2}}\, p^{-z}}\,,
\hspace{10mm}
\text{for }p>2\,.
\nonumber
\end{align}
In analogy with \eqref{AsuperReal}, Ref. \cite{ruelle1989adelic} proposed a $p$-adic superamplitude:
\begin{align}
\mathcal{A}_4^{(p)}\big(s,t\big)=\Gamma^{(-1)}_{p}(-s)\,\Gamma^{(-1)}_{p}(-t)\,\Gamma^{(-1)}_{p}(s+t)\,.
\label{AsuperPadic}
\end{align}
Unlike the $p$-adic Veneziano amplitude and the alternative $p$-adic superstring amplitudes studied in \cite{aref1988p,marshakov1990new,garcia2022towards}, there is no integral formula to justify the identification of \eqref{AsuperPadic} with a string theory amplitude. The motivation for \eqref{AsuperPadic} was the goal of obtaining an adelic formula for superstring amplitudes. The signed $p$-adic zeta function satisfies the product formula
\begin{align}
\prod_{p>2}\zeta_p^{(-1)}(z)=\prod_{p>2} \frac{1}{1+(-1)^{\frac{p+1}{2}}\,p^{-z}}=L_{4,2}(z)\equiv\sum_{n=1}^\infty \frac{\chi_{4,2}(n)}{n^z}\,, \label{prodZetaSigned}
\end{align}
where the product over primes can be shown to converge absolutely for Re$[z]>1$ and conditionally for Re$[z]>\frac{1}{2}$. Here $L_{4,2}(z)$ is the Dirichlet L-function for the field of Gaussian rationals, and $\chi_{4,2}(n)$ is the Dirichlet character with modulus 4 and index 2, whose value for any integer argument can be determined from the facts that $\chi_{4,2}(n_1 n_2)=\chi_{4,2}(n_1) \chi_{4,2}(n_2)$ for any $n_1,n_2\in \mathbb{N}$, $\chi_{4,2}(2)=0$, and $\chi_{4,2}(p)=(-1)^{\frac{p+3}{2}}$ for any prime $p>2$. By virtue of being a Dirichlet L-function, $L_{4,2}(z)$ obeys a functional equation:
\begin{align}
L_{4,2}(1-z)
=\frac{4^{z}}{2i}\,\Gamma_\infty^{(-1)}(z)\,
L_{4,2}(z)\,.
\label{superFunctional}
\end{align}
Using this relation, the real superamplitude $\mathcal{A}_4^{(\infty)}$ can be recast solely in terms of the Dirichelet L-function: 
\begin{align}
\mathcal{A}_4^{(\infty)}
=-8\,\frac{L_{4,2}(1+s)L_{4,2}(1+t)L_{4,2}(1+u)}{L_{4,2}(-s)L_{4,2}(-t)L_{4,2}(-u)}\,.
\end{align}
Also through the use of the functional equation, we observe that if we set aside the criterion Re$[z]>\frac{1}{2}$ required for the convergence of \eqref{prodZetaSigned} and allow ourselves to split up an infinite product, then we find that
\begin{align}
\prod_{p\in\mathbb{P}} \Gamma_p^{(-1)}(z)
=
\Gamma_2^{(-1)}(z)
\frac{\prod_{p>2}\zeta_p^{(-1)}(z)}{\prod_{p>2}\zeta_p^{(-1)}(1-z)}
=i\,2^{2z-1}\frac{L_{4,2}(z)}{L_{4,2}(1-z)}
=-\frac{1}{\Gamma_\infty^{(-1)}(z)}\,.
\label{gammaSignedProd}
\end{align}
 If we are also allowed to split an infinite product over $\mathcal{A}_4^{(p)}$ into three products over the signed gamma function in \eqref{AsuperPadic}, then, as in  \cite{ruelle1989adelic}, we arrive at the following formula: 
\begin{align}
\mathcal{A}_4^{(\infty)}(s,t)
\prod_{p \in \mathbb{P}}
\mathcal{A}_4^{(p)}(s,t)
=-1
\,.
\label{superAdelicProd}
\end{align}

\subsection{Prefactor regulated adelic superamplitude}
\label{4.1}
\begin{figure}[t]
\begin{center}
\includegraphics[width=0.4\linewidth]{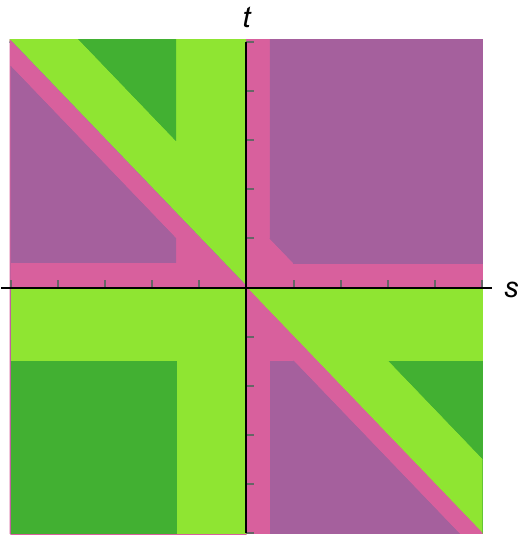}
\end{center}
\caption{Different regions of kinematic space for 4-particle gluon scattering. The regions in {\bf \textcolor{colour5}{light green}} and {\bf \textcolor{colour6}{dark green}} represent physical regions, while the {\bf \textcolor{colour7}{magenta}} and {\bf \textcolor{colour8}{purple}} regions are unphysical. In the {\bf \textcolor{colour6}{dark green}} regions, the product for the adelic amplitude can be rendered convergent by dressing the $p$-adic amplitudes with coefficients $\mathcal{C}_p=(-1)^{\frac{p+1}{2}}p^2$. In the {\bf \textcolor{colour8}{purple}} regions, convergence is achieved with coefficients $\mathcal{C}_p=(-1)^{\frac{p+1}{2}}p$.
}
\label{fig:SuperRegions}
\end{figure}

In the interest of finding an alternative to the constant adelic superamplitude not requiring the assumptions needed to arrive at \eqref{superAdelicProd}, we introduce coefficients $\mathcal{C}_p$ and take the adelic amplitude to be given by
\begin{align}
\mathcal{A}_4^{(\mathbb{A})}(s,t)
=
\mathcal{A}_4^{(\infty)}(s,t)
\prod_{p \in \mathbb{P}}
\mathcal{C}_p\, \mathcal{A}_4^{(p)}(s,t)
\,.
\label{AadelSuper}
\end{align}
If there exists a theory to which such a product can be meaningfully associated, then it will be necessary to multiply $\mathcal{A}_4^{(\mathbb{A})}$ with a kinematic factor like $K(k_i,\zeta_i)$ in equation \eqref{Agluon}, but we will not examine this complication. As in the bosonic case, there exist two choices of $\mathcal{C}_p$ for which the product in \eqref{AadelSuper} converges in particular kinematic regimes, but there is no choice of $\mathcal{C}_p$ that will render the product convergent throughout the physical regimes given by
\begin{align}
\nonumber
&    \text{$s$-channel:} \hspace{3mm} s\geq 0\,, \hspace{5mm} -s\leq t \leq 0\,,
    \\[3pt]
    \label{superSubsets}
&    \text{$t$-channel:} \hspace{3.5mm} t\geq 0\,, \hspace{5.5mm} -t\leq s \leq 0\,,
    \\[3pt]
&    \text{$u$-channel:} \hspace{3mm} s,t<0\,,
\nonumber
\end{align}
and depicted in different shades of green in figure \ref{fig:SuperRegions}. If we set $\mathcal{C}_p=(-1)^{\frac{p+1}{2}}p^2$ for $p>2$, convergence is attained in the regions marked in {\bf \textcolor{colour6}{dark green}} in figure \ref{fig:SuperRegions} and given by
\begin{align*}
&    \text{$s$-channel subset:} \hspace{3mm} t < -\frac{3}{2}\,, \hspace{5mm} s> \frac{3}{2} - t\,,
    \\[3pt]
&    \text{$t$-channel subset:} \hspace{3mm} s < -\frac{3}{2}\,, \hspace{5mm} t> \frac{3}{2} - s\,,
    \\[3pt]
&    \text{$u$-channel subset:} \hspace{3mm} s,t<-\frac{3}{2}\,.
\end{align*}
And if we set $\mathcal{C}_p=(-1)^{\frac{p+1}{2}}p$  for $p>2$, the product is convergent for
\begin{align*}
&   \mathcal{A}) \hspace{3mm} s+t<-\frac{1}{2}\,, \hspace{5.5mm}  t>\frac{1}{2}\,,   \hspace{5.5mm} s< -\frac{3}{2}\,,
    \\[3pt]
&   \mathcal{B}) \hspace{3mm} s+t<-\frac{1}{2}\,, \hspace{5.5mm}  s>\frac{1}{2}\,,   \hspace{5.5mm} t< -\frac{3}{2}\,,
    \\[3pt]
&   \mathcal{C}) \hspace{3mm} s,t>\frac{1}{2}\,,  \hspace{5.5mm} s+t> \frac{3}{2}\,.
\end{align*}
These regions are depicted in {\bf \textcolor{colour8}{purple}} in  figure \ref{fig:SuperRegions}. For $p=2$, we set $\mathcal{C}_2=4i$ and $\mathcal{C}_2=2i$ respectively, but the precise values of $\mathcal{C}_2$ are unimportant since changing a finite number of coefficients only affects the overall normalization of $\mathcal{A}^{(\mathbb{A})}_4$, which we do not determine.

For the values of $\mathcal{C}_p$ just described, it is a simple exercise to evaluate the product \eqref{AadelSuper} in the convergent regions through the use of the formula
\begin{align}
(-1)^{\frac{p+1}{2}}p^n\frac{\zeta^{(-1)}_p(a+b-n)}{\zeta^{(-1)}_p(a)\,\zeta^{(-1)}_p(b)}
=
\frac{\zeta_p^{(-1)}(-a-b+n)}{\zeta^{(-1)}_p(-a)\,\zeta^{(-1)}_p(-b)}\,.
\end{align}
Analytically continuing the products obtained for $\mathcal{C}_p=(-1)^\frac{p+1}{2}p^2$ in the convergent regimes \eqref{superSubsets} to the full respective scattering channels, and continuing for $\mathcal{C}_p=(-1)^\frac{p+1}{2}p$ the product in region $\mathcal{A})$ to the $s$-channel, in $\mathcal{B})$ to the $t$-channel, and in $\mathcal{C})$ to the $u$-channel, we get the following candidate amplitudes:
\begin{align}
\mathcal{C}_p=(-1)^{\frac{p+1}{2}}p^2:
\hspace{5mm}
\mathcal{A}^{(\mathbb{A})}_4=
\begin{cases}
\displaystyle 
\,-4\,\frac{L_{4,2}(s)L_{4,2}(1+t)L_{4,2}(1+u)}{L_{4,2}(-s)L_{4,2}(-1-t)L_{4,2}(-1-u)}
&\hspace{5mm}\text{ $s$-channel}\,,
\\
\\
\displaystyle 
\,-4\,\frac{L_{4,2}(1+s)L_{4,2}(t)L_{4,2}(1+u)}{L_{4,2}(-1-s)L_{4,2}(-t)L_{4,2}(-1-u)}
&\hspace{5mm}\text{ $t$-channel}\,,
\\
\\
\displaystyle 
\,-4\,\frac{L_{4,2}(1+s)L_{4,2}(1+t)L_{4,2}(u)}{L_{4,2}(-1-s)L_{4,2}(-1-t)L_{4,2}(-u)}
&\hspace{5mm}\text{ $u$-channel}\,,
\end{cases}
\label{newAdelicSuper}
\end{align}

\begin{align}
\mathcal{C}_p=(-1)^{\frac{p+1}{2}}p:
\hspace{5mm}
\mathcal{A}^{(\mathbb{A})}_4=
\begin{cases}
\displaystyle 
\,-2\,\frac{L_{4,2}(1+s)L_{4,2}(t)L_{4,2}(u)}{L_{4,2}(-1-s)L_{4,2}(-t)L_{4,2}(-u)}
&\hspace{5mm}\text{ $s$-channel}\,,
\\
\\
\displaystyle
\,-2\,\frac{L_{4,2}(s)L_{4,2}(1+t)L_{4,2}(u)}{L_{4,2}(-s)L_{4,2}(-1-t)L_{4,2}(-u)}
&\hspace{5mm}\text{ $t$-channel}\,,
\\
\\
\displaystyle 
\,-2\,\frac{L_{4,2}(s)L_{4,2}(t)L_{4,2}(1+u)}{L_{4,2}(-s)L_{4,2}(-t)L_{4,2}(-1-u)}
&\hspace{5mm}\text{ $u$-channel}\,.
\end{cases}
\end{align}
In the next two subsections, we will analyze the partial wave decompositions of these expressions and study their high energy asymptotics.

\subsection{Partial Wave Decomposition}
\label{4.2}
For massless 4-particle scattering in $d$ spacetime dimensions, the partial wave decomposition reads
\begin{align}
\underset{s=s^\ast}{\text{Res}}\,\mathcal{A}\Big(s,\frac{s(\cos\theta-1)}{2}\Big)
=\sum_{m\in \mathbb{N}_0}K^{(s^\ast)}_m(d)\,C_m^{(\frac{d-3}{2})}(\cos\theta)\,,
\label{GegSup}
\end{align}
where unitarity stipulates that all the coefficients $K_m^{(s^\ast)}$ are non-negative. For the real superamplitude $\mathcal{A}_4^{(\infty)}(s,t)$ the $s$-channel poles are situated at $s=s^\ast\in 2\mathbb{N}_0+1$, and the number of Gegenbauer polynomials with non-zero weight in the expansion \eqref{Geg} grows linearly with $s^\ast$ , being given by $(s^\ast+1)/2$ in generic spacetime dimensions. The first residue is simply a constant:
\begin{align}
&\underset{s=1}{\text{Res}}\,\mathcal{A}_4^{(\infty)}\Big(s,\frac{(s+4)(\cos\theta-1)}{2}\Big)=\,
4\pi=4\pi\,C_0^{(\frac{d-3}{2})}(\cos\theta)\,.
\end{align}
The second residue decomposes as
\begin{align}
\underset{s=3}{\text{Res}}\,\mathcal{A}_4^{(\infty)}\Big(s,\frac{(s+4)(\cos\theta-1)}{2}\Big)
=\,&\frac{\pi}{6}(9\cos^2\theta-1)
\\[6pt]
=\,&
\frac{(10-d)\pi}{6(d-1)}\,
C_0^{(\frac{d-3}{2})}(\cos\theta)
+
\frac{3\pi}{d^2-4d+3}\,
C_2^{(\frac{d-3}{2})}(\cos\theta)\,.
\nonumber
\end{align}
From the coefficient $K_0^{(3)}(d)$, we see that the real superamplitude violates unitarity for $d>10$, in accordance with 10 being the critical dimension of superstring theory. 

\begin{figure}
\begin{center}
\includegraphics[width=0.99\linewidth]{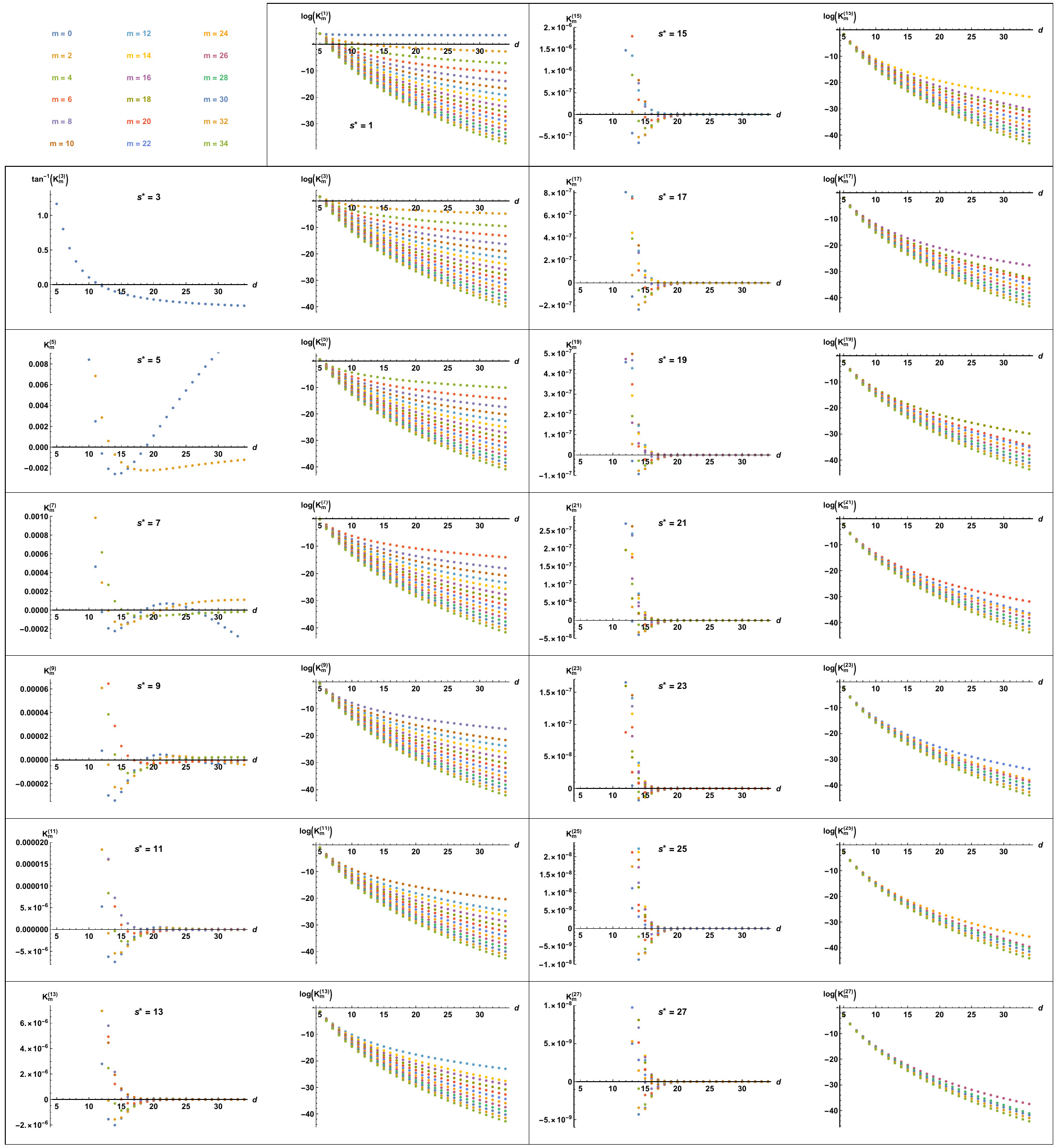}
\end{center}
\caption{Coefficients in the $s$-channel Gegenbauer decomposition of the tentative adelic superamplitude with $\mathcal{C}_p=(-1)^{\frac{p+1}{2}}p^2$. In five to 11 dimensions, all the coefficients are positive. In 12 dimensions and higher, some coefficients are negative. For $d\leq 4$, the coefficients blow up. For the sake of visibility, the logarithm has been taken of coefficients $K^{(s^\ast)}_{m}(d)$ that are positive for all $d>4$.
}
\label{fig:coeffsSuper}
\end{figure}

\begin{figure}
\begin{center}
\includegraphics[width=0.99\linewidth]{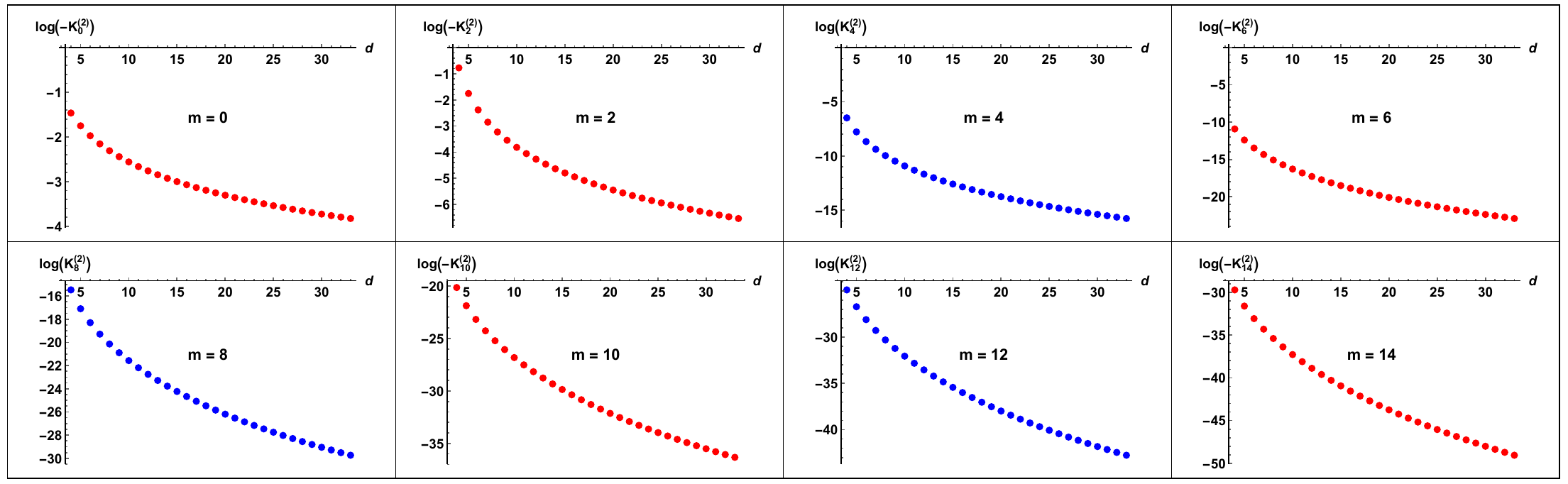}
\end{center}
\caption{Coefficients $K_m^{(2)}(d)$ in the $s$-channel Gegenbauer decomposition of the tentative adelic superamplitude with $\mathcal{C}_p=(-1)^{\frac{p+1}{2}}p$ for the residue at $s=2$. The coefficients $K_m^{(2)}(d)$ are positive for all $d>3$ when $m=4$, $8$ or $12$, marked in \textcolor{blue}{\bf{blue}}, and are all negative for $m=0$, $2$, $6$, $10$, and $14$, marked in marked in \textcolor{red}{\bf{red}}, which implies unitarity violations. For $d\leq 3$, the coefficients blow up. For the sake of visibility, $\log|K_m^{(2)}|$ rather than $K_m^{(2)}$ is plotted along the $y$-axis.}
\label{fig:coeffsSuper2}
\end{figure}
\subsubsection*{$\mathcal{C}_p=(-1)^{\frac{p+1}{2}}p^2$ adelic amplitude}

$L_{4,2}(x)$ equals zero whenever $x\in -2\mathbb{N}_0-1$, and as a result the $s$-channel poles of the $\mathcal{C}_p=(-1)^{\frac{p+1}{2}}p^2$ adelic amplitude are located at the values $s=s^\ast \in 2\mathbb{N}_0+1$ as in the real case. But unlike the real case, every even Gegenbauer polynomial is present at every residue. For the first 14 residues, figure \ref{fig:coeffsSuper} depicts the first 18 Gegenbauer coefficients, computed numerically using the orthogonality relation \eqref{GegOrtho}. On the boundary of the physical region, there is a pole at $t=0$. Wherever the line $t=0$ intersects any of the lines $s=s^\ast$, there is a double pole, causing the decomposition \eqref{GegSup} to blow up for $d\leq 4$. For the first pole at $s^\ast=1$, the coefficients are all positive for every $d>4$, but at higher values of $s^\ast$ an increasing number of coefficients $K_m^{(s^\ast)}(d)$ are negative for some values of $d$. The first unitarity violation occurs between $d=11$ and $d=12$ for the pole at $s^\ast=3$. In dimensions 5 to 11, all the computed coefficients are positive.

\subsubsection*{$\mathcal{C}_p=(-1)^{\frac{p+1}{2}}p$ adelic amplitude}

The $\mathcal{C}_p=(-1)^{\frac{p+1}{2}}p$ adelic amplitude has $s$-channel poles situated at $s^\ast \in 2\mathbb{N}_0$. In the physical $s$-channel region, there are no poles in $t$ and $u$, and therefore no double poles at the intersection of $s$ and $t$ poles, in consequence whereof we can numerically compute the Gegenbauer coefficients for $d > 3$. The residue at $s=0$ is a constant given by
\begin{align}
\underset{s=1}{\text{Res}}\,\mathcal{A}_4^{(\mathbb{A})}\Big(s,\frac{(s+4)(\cos\theta-1)}{2}\Big)
=\frac{\pi}{2L'_{4,2}(-1)}\approx 2.69377\,.
\end{align}
While this number is positive, it is evident from the plots of Gegenbauer coefficients for the residue at $s=2$ shown in figure \ref{fig:coeffsSuper2} that other coefficients are negative. Unitarity is violated in every number of dimensions. 

\subsection{High energy limit}
\label{4.3}
From the asymptotic behaviour of the gamma function \eqref{gammaAsymptotic}, it can be shown that the real superamplitude has the following high energy limits:
\begin{align}
&
\text{large $s$, fixed $t$:} \hspace{12mm}
\mathcal{A}_4^{(\infty)}(s,t) \approx
4\pi\, s^{t-1}\,
\Gamma(-t)
\sec\big(\frac{\pi s}{2}\big)
\sin\big(\frac{\pi t}{2}\big)
\sin\big(\frac{\pi (s+t)}{2}\big)\,,
\\[6pt]
&\text{large $s$, fixed $\theta$:} \hspace{12mm}
\mathcal{A}_4^{(\infty)}\Big(s,\frac{s}{2}(\cos\theta-1)\Big)\approx
\\[4pt]
&
\hspace{20mm}
\bigg|\frac{2}{\sin\theta\cot\big(\frac{\theta}{2}\big)^{\cos\theta}}\bigg|^{-s}
\,
\sqrt{\frac{32\,\pi^3}{\sin^2\theta\,s^3}}
\,
\bigg(
1-\cos\Big(\frac{\pi s\cos\theta}{2}\Big)
\sec\big(\frac{\pi s}{2}\big)
\bigg)\,. \label{superRegge}
\end{align}
We see that $\mathcal{A}_4^{(\infty)}(s,t)$ scales as $s^{t-1}$ in the Regge limit and decays exponentially in the high energy fixed scattering angle limit. To see that this benign high energy behaviour carries over in the adelic case, we express the relationship between the real and adelic amplitudes as
\begin{align}
\mathcal{A}_4^{(\mathbb{A})}\big|_{s\text{-channel}}(s,t)=\mathcal{A}_4^{(\infty)}(s,t)\,\mathcal{A}_4^{(\mathbb{P})}(s,t)\big|_{s\text{-channel}}\,,
\end{align}
where the factor relating the two is given by
\begin{align}
&
\mathcal{C}_p=(-1)^{\frac{p+1}{2}}p^2:
\hspace{8mm}
\mathcal{A}^{(\mathbb{P})}_4(s,t)\big|_{s\text{-channel}}
=
 \frac{L_{4,2}(s)\,L_{4,2}(-t)\,L_{4,2}(s+t)}{2\,L_{4,2}(1+s)\,L_{4,2}(-1-t)\,L_{4,2}(-1+s+t)}
\,,
\nonumber
\\
&
\\[-6pt]
&
\mathcal{C}_p=(-1)^{\frac{p+1}{2}}p:
\hspace{10mm}
\mathcal{A}^{(\mathbb{P})}_4(s,t)\big|_{s\text{-channel}}
=
 \frac{L_{4,2}(-s)\,L_{4,2}(t)\,L_{4,2}(-s-t)}{4\,L_{4,2}(-1-s)\,L_{4,2}(1+t)\,L_{4,2}(1-s-t)}
\,.
\nonumber
\end{align}
Since $L_{4,2}(x)$ tends to one as $x\rightarrow \infty$, we immediately infer that the $\mathcal{C}_p=(-1)^{\frac{p+1}{2}}p^2$ adelic amplitude has the same high energy asymptotics as the real amplitude. In the case $\mathcal{C}_p=(-1)^{\frac{p+1}{2}}p$, it follows from the functional equation \eqref{superFunctional} that
\begin{align}
&\mathcal{C}_p=(-1)^{\frac{p+1}{2}}p:
\hspace{35mm}
\mathcal{A}^{(\mathbb{P})}_4(s,t)\big|_{s\text{-channel}}=
\label{100}
\\[6pt]
&\hspace{14mm}
-\frac{t(s+t)
\cot\big(\frac{\pi s}{2}\big)
\cot\big(\frac{\pi t}{2}\big)
\cot\big(\frac{\pi (s+t)}{2}\big)
}{2\pi (1+s)} \frac{L_{4,2}(1+s)\,L_{4,2}(1-t)\,L_{4,2}(1+s+t)}{4\,L_{4,2}(2+s)\,L_{4,2}(-t)\,L_{4,2}(s+t)}
\,.
\nonumber
\end{align}
In the limit of large $s$ and fixed $t$, this expression tends to a periodic function, and in the limit of large $s$ and fixed $\theta$, \eqref{100} tends to a periodic function in $s$ times a linearly growing function, which is suppressed compared to the exponential decay of \eqref{superRegge}.

\section{Discussion}
\label{5}
We have seen that the adelic 5-point amplitude, given by a sometimes convergent product over real and $p$-adic amplitudes, admits exact evaluation in special kinematic regimes and that the results of these evaluations suggest seeking out alternative regularization procedures for interpreting the 4-point amplitude, which is always given by a divergent product. We analyzed in detailed a particular procedure that consists in endowing the $p$-adic amplitudes with momentum-independent coefficients so as to render their product convergent. Similarly, the 5-point and higher $p$-adic amplitudes can be dressed with coefficients that drastically alter their products and modify their regimes of convergence, and imposing factorization may point to a consistent set of choices for these coefficients, although it would be preferable to derive them from first principles.

Of the four non-constant adelic 4-point amplitudes we obtained through regularization via coefficients, three contain double poles at the boundary of the physical kinematic region but appear consistent with unitary, while the fourth contains no double poles in the physical region but violates unitarity. And all four adelic amplitudes exhibit non-polynomial residues, resulting in the presence of an infinite tower of Gegenbauer polynomials---a peculiarity which, incidentally, was also present in the Coon amplitude in the form in which it was originally written down. Only recently did Figueroa and Tourkine \cite{figueroa2022unitarity} remedy this situation by multiplying the Coon amplitude with an extra function of the Mandelstam invariants which made the residues polynomial and enabled them to provide compelling evidence for the unitarity of the corrected Coon amplitude.

The absence of crossing symmetry in the adelic amplitudes is certainly an unusual feature but not one that need be fatal be to the existence of an underlying theory. It has been known since the work of Bros, Epstein, and Glaser \cite{bros1964some,bros1965proof} that locality, causality, and unitarity together imply crossing symmetry in theories with a mass gap, but for theories with massless particles, if we abandon the mathematical assumption of analyticity, no such proof is known.\footnote{We refer the reader to Mizera's paper \cite{mizera2021crossing} for a nice review of, and recent progress on, crossing symmetry.} Taking a different view on piecewise analyticity and lack of crossing symmetry, the adelic formalism provides a method of engineering features otherwise difficult to realize for tree-amplitudes in string theory and quantum field theory, where Feynman tree-diagrams are largely insensitive to whether external momenta are in- or out-going. As we have seen, these conditions are crucial to the evaluation of infinite products with disjoint regions of convergence.

The most intriguing feature of the adelic amplitudes in \eqref{newAdelic1}, \eqref{newAdelic2}, and \eqref{newAdelicSuper} is arguably the numerical evidence of unitarity for target spaces of suitable dimensionalities. Unitarity of string theory tree-amplitudes has previously proved itself to be a powerful principle capable of deriving new mathematical results, as in the work of Green and Wen \cite{green2019superstring}. For the candidate amplitude \eqref{newAdelic1}, $d=27$ is the largest number of dimensions before unitarity violations are observed, while for the tentative amplitudes in  \eqref{newAdelic2} and \eqref{newAdelicSuper} these numbers are $d=10$ and $d=11$ respectively, although \eqref{newAdelic2} should perhaps be dismissed, since it suffers from the malady of exchanging higher spins on a massless pole. It remains an intriguing prospect to ascertain if any of the above numbers truly represents the critical dimension of the adelic string. It will also be potentially interesting to investigate whether there is evidence for unitarity of more broad classes of adelic amplitudes associated to general Dirichlet L-functions, and if so, to determine the associated critical dimensions. The family of local gamma functions in \eqref{signedGammas}, indexed by a superscript of minus one, admit of a generalization to families of gamma functions indexed by any non-zero rational number, and for each of these families it is possible to form an adelic construction. It would also be desirable to establish partial wave unitarity, if this is a genuine feature, on a firmer footing than a finite number of numerical tests. For unitary to hold, a doubly infinite set of constraints must be obeyed: for each residue at an infinite tower of poles, each coefficient in an infinite sum of Gegnebauer polynomials must be non-negative. The fact that these conditions are all satisfied in the first many instances, where they have so far been checked, is a phenomenon that is difficult to explain without positing an underlying unitary theory associated to the ring of adeles.

\section*{Acknowledgements}
I am indebted to Gabriel Cuomo, Matthew Heydeman, An Huang, Ziming Ji, Justin Kaidi, Joseph Minahan, Yaron Oz, Gabi Zafrir, and Wayne Zhao for illuminating discussions.

\appendix

\section{Regulating $S_1(z\,|\,\omega)$ with Coefficients}
\label{A}

Divergent infinite products occur in many instances in physics outside the context of adelic amplitudes. One particular class of examples are furnished by computations of partition functions of superconformal theories in 5$d$ \cite{lockhart2018superconformal,imamura2013perturbative,minahan20135d} through the appearance of the multiple sine function $S_r(z\,|\,\vec{\omega})$. This function carries a subscript $r\in\mathbb{N}$ and depends on a complex argument $z$ as well as on an $r$-tuple  $\vec{\omega}=(\omega_1,...,\omega_r)$ of complex parameters all with positive real parts. Informally, the multiple sine function can be expressed as an infinite product:
\begin{align}
S_r(z\,|\,\vec{\omega}) \sim
\prod_{n_1,...,n_r=0}^\infty
\Big(\vec{n}\cdot \vec{\omega}+\sum_{i=1}^r\omega_i-z\Big)\big(\vec{n}\cdot \vec{\omega}+z\big)^{(-1)^{r+1}}\,,
\label{Sr}
\end{align}
which needs to be regulated in order to extract a finite value. A more formal definition of the function can be formulated by first introducing the Barnes zeta function, given for complex $s$ with Re$[s]>r$ by
\begin{align}
\zeta_r(z,s\,|\,\vec{\omega})
=\sum_{n_1,...,n_r=0}^\infty
\Big(\vec{n}\cdot\vec{\omega}+z\Big)^{-s}\,,
\end{align}
and defined by analytic continuation for all other $s\in \mathbb{C}$ except for single poles at $s=1,2,..., r$. Invoking the Barnes zeta function, the multiple gamma function can be defined as
\begin{align}
\Gamma_r(z\,|\,\vec{\omega})
=\exp\bigg(
\frac{\partial}{\partial s}
\zeta_r(z,s\,|\,\vec{\omega})\Big|_{s=0}
\bigg)\,.
\label{Gammar}
\end{align}
With these definitions in place, the multiple sine function is defined as
\begin{align}
S_r(z|\omega)
=\frac{\Gamma_r\Big(\sum_{i=1}^r \omega_i-z\,\Big|\,\vec{\omega}\Big)^{(-1)^r}}
{\Gamma_r(z\,|\,\vec{\omega})}\,.
\label{SrFormal}
\end{align}

In this appendix, we demonstrate how the regularization procedure of subsection \ref{3.1} applied to the divergent product \eqref{Sr} recovers the formal definition \eqref{SrFormal} of the multiple sine function, restricting attention to the simplest case $r=1$, for which $S_r$ is given by
\begin{align}
S_1(z\,|\,\omega) \sim \prod_{n=0}^\infty (n \omega +\omega - z)(n \omega +z)\,.
\label{S1}
\end{align}
To determine the $z$-dependence of this function at fixed $\omega$ our strategy is to rescale the individual factors by $z$-independent coefficients so as to render the product convergent. At large $n$ the factors in \eqref{S1} scale as $n^2\omega^2$, so as a first step we can divide all factors with $n>0$ by this quantity:
\begin{align}
S_1(z\,|\,\omega) \sim  (\omega-z)z\prod_{n=1}^\infty \Big(1 + \frac{\omega - z}{n \omega}\Big)\Big(1 +\frac{z}{n\omega}\Big)\,.
\end{align}
At large $n$, the factors of the infinite product now tend to one, but they scale as $\big(1+\frac{1}{n}+\mathcal{O}(\frac{1}{n^2})\big)$, which still leaves the product divergent. To achieve convergence we must divide by a quantity with the same large $n$ scaling. We could simply divide by $\big(1+\frac{1}{n}\big)$, but it will be more convenient for us to make the choice $e^{1/n}$. We arrive then at the regulated answer for the single sine function
\begin{align}
S_1(z\,|\,\omega) = C(\omega)\, (\omega-z)z\prod_{n=1}^\infty e^{-1/n}\Big(1 + \frac{\omega - z}{n \omega}\Big)\Big(1 +\frac{z}{n\omega}\Big)\,,
\label{S1regulated}
\end{align}
where $C(\omega)$ is an unknown $z$-independent quantity. Dividing by $\big(1+\frac{1}{n}\big)$ instead of $e^{1/n}$ would have amounted to a redefinition of $C(\omega)$.

The multiple gamma function \eqref{Gammar} with $r=1$ is known to equal
\begin{align}
\Gamma_1(z \,|\, \omega)
= \frac{\omega^{\frac{z}{\omega}-\frac{1}{2}}}{\sqrt{2\pi}}\,\Gamma\big(\frac{z}{\omega}\big)\,. 
\end{align}
According to the formal definition of the multiple sine function \eqref{SrFormal}, this means the single sine function is given by 
\begin{align}
S_1(z|\omega)
=\frac{1}
{\Gamma_1(\omega-z\,|\,\omega)\Gamma_1(z\,|\,\omega)}
=\frac{2\pi}{\Gamma(\frac{z}{\omega})\,\Gamma(\frac{\omega-z}{\omega})}\,.
\label{S1exact}
\end{align}
By twice applying Weierstrass' definition of the gamma function
\begin{align}
\Gamma(z)=\frac{e^{-\gamma z}}{z}\prod_{n=1}^\infty \frac{e^{z/n}}{1+\frac{z}{n}}\,,
\end{align}
where $\gamma$ is the Euler-Mascheroni constant, we see that \eqref{S1exact} agrees with the result \eqref{S1regulated} of the regulated infinite product, and we determine the $\omega$-dependent normalization in \eqref{S1regulated} to be given by
\begin{align}
C(\omega) = 2\pi\frac{e^\gamma}{\omega^2}\,.
\end{align}

\bibliographystyle{ssg}
\bibliography{adelic}

\end{document}